\documentclass[traditabstract]{aa}
\usepackage{graphicx}
\usepackage{natbib}
\usepackage{amssymb}
\bibpunct{(}{)}{;}{a}{}{,} 

\usepackage{longtable}
\usepackage{url}
\usepackage{lscape}

\newcommand{\beqa}{\begin{eqnarray}} 
\newcommand{\eeqa}{\end{eqnarray}}

\newcommand{\bsub}{\begin{subequations}}
\newcommand{\esub}{\end{subequations}}
\newcommand{\beal}{\begin{align}}
\newcommand{\ealn}{\end{align}}

\begin{document}

\title{Supernova spectra below strong circumstellar interaction}
\titlerunning{Supernova spectra below strong circumstellar interaction}

\author{
G.~Leloudas\inst{1,2,3}
\and  E.~Y.~Hsiao\inst{\ref{inst4},\ref{inst5}}
\and J.~Johansson\inst{1}
\and K.~Maeda\inst{\ref{instXTRA},\ref{inst6}}
\and T.~J.~Moriya\inst{\ref{inst7}}
\and J.~Nordin\inst{\ref{inst8},\ref{inst9}}
\and T.~Petrushevska\inst{1}
\and J.~M.~Silverman\inst{\ref{inst10}}
\and J.~Sollerman\inst{\ref{inst11}}
\and M.~D.~Stritzinger\inst{\ref{inst5}}
\and F.~Taddia\inst{\ref{inst11}}
\and D.~Xu\inst{3}
}

\institute{ 
The Oskar Klein Centre, Department of Physics, Stockholm University, AlbaNova, 10691 Stockholm, Sweden \label{inst1}
\\ \email{giorgos@dark-cosmology.dk}
\and Department of Particle Physics \& Astrophysics, Weizmann Institute of Science, Rehovot 76100, Israel \label{inst2}
\and Dark Cosmology Centre, Niels Bohr Institute, University of Copenhagen, Juliane Maries Vej 30, 2100 Copenhagen, Denmark \label{inst3}
\and Carnegie Observatories, Las Campanas Observatory, Colina El Pino, Casilla 601, Chile \label{inst4}
\and Department of Physics and Astronomy, Aarhus University, Ny Munkegade 120, 8000 Aarhus C, Denmark \label{inst5}
\and Department of Astronomy, Kyoto University, Kitashirakawa-Oiwake-cho, Sakyo-ku, Kyoto 606-8502, Japan \label{instXTRA}
\and Kavli Institute for the Physics and Mathematics of the Universe (WPI), Todai Institutes for Advanced Study, University of Tokyo, Kashiwanoha 5-1-5, Kashiwa, Chiba 277-8583, Japan \label{inst6}
\and Argelander Institute for Astronomy, University of Bonn, Auf dem HŸgel 71, D-53121 Bonn, Germany \label{inst7}
\and Space Sciences Laboratory, University of California Berkeley, Berkeley, CA 94720, USA  \label{inst8}
\and E. O. Lawrence Berkeley National Lab, 1 Cyclotron Rd., Berkeley, CA 94720, USA  \label{inst9}
\and Department of Astronomy, University of Texas, Austin, TX 78712-0259, USA  \label{inst10}
\and The Oskar Klein Centre, Department of Astronomy, Stockholm University, AlbaNova, 10691 Stockholm, Sweden \label{inst11}
}

\date{Received -- / Accepted --}

\abstract
{
We construct spectra of supernovae (SNe) interacting strongly with a circumstellar medium (CSM) by adding SN templates, a black-body continuum, and an emission-line spectrum.
In a Monte Carlo simulation we vary a large number of parameters, such as the SN type, brightness and phase, the strength of the CSM interaction, the extinction, and the signal to noise (S/N) of the observed spectrum.
We generate more than 800 spectra, distribute them to ten different human classifiers, and study how the different simulation parameters affect the appearance of the spectra and their classification.
The SNe~IIn showing some structure over the continuum were characterized as `SNe~IInS' to allow for a better quantification. 
We demonstrate that the flux ratio of the underlying SN to the continuum  $f_V$ is  the single most important parameter determining whether a spectrum can be classified correctly.
Other parameters, such as extinction, S/N, and the width and strength of the emission lines, do not play a significant role.
Thermonuclear SNe get progressively classified as Ia-CSM, IInS, and IIn as $f_V$ decreases.
The transition between Ia-CSM and IInS occurs at $f_V \sim 0.2-0.3$.
It is therefore possible to determine that SNe~Ia-CSM are found at the (un-extincted) magnitude range $-19.5 > M > -21.6$, in very good agreement with observations, and that the faintest SN~IIn that can hide a SN~Ia has $M = -20.1$.
The literature sample of SNe~Ia-CSM shows an association with 91T-like SNe~Ia. 
Our experiment does not support that this association can be attributed to a luminosity bias (91T-like being brighter than normal events).
We therefore conclude that this association has real physical origins and we propose that 91T-like explosions  result from single degenerate progenitors that are responsible for the CSM.
Despite the spectroscopic similarities between SNe~Ibc and SNe~Ia, the number of misclassifications between these types was very small in our simulation and mostly at low S/N. Combined with the SN luminosity function needed to reproduce the observed SN~Ia-CSM luminosities, it is unlikely that  SNe~Ibc constitute an important contaminant within this sample.
We show how Type~II spectra transition to IIn and how the H$\alpha$ profiles vary with $f_V$.
SNe~IIn fainter than $M = -17.2$ are unable to mask SNe~IIP brighter than $M = -15$.
A more advanced simulation, including radiative transfer, shows that our simplified model is a good first order approximation. 
The spectra obtained are in good agreement with real data.
}

\keywords{supernovae: general}

\maketitle

\section{Introduction}

Type~IIn supernovae \citep[SNe~IIn;][]{1990MNRAS.244..269S} are distinguished by the narrow lines in their spectra \citep{1997ARA&A..35..309F}. The power source behind SN~IIn luminosities,  at the same time giving rise to their spectral appearance, is the interaction between fast moving ejecta and circumstellar material (CSM),  ejected from the progenitor system at earlier stages \citep[e.g.][]{1994MNRAS.268..173C}. 
The IIn classification is purely phenomenological and does not account for the physical nature of the explosion (thermonuclear or core collapse) or the eruption (i.e. without destruction of the stellar progenitor) hidden below the CSM interaction. 

In at least two cases, SNe~IIn have been observationally connected to  the explosion of a very massive star
\citep{2009Natur.458..865G,2011ApJ...732...63S}, leading to the conclusion that a fraction of these explosions are unambiguously of core-collapse nature. 
SN impostors are objects that were initially misclassified as SNe~IIn because of their spectral appearance.
These objects have been linked to the eruptions of massive luminous blue variable (LBV) stars \citep{2000PASP..112.1532V,2006MNRAS.369..390M}.
SN~2009ip was a SN impostor that might have exploded as a real SN \citep{2013MNRAS.430.1801M}, although this is a topic of debate 
\citep[e.g.][]{2013ApJ...767....1P,2013MNRAS.433.1312F,2014ApJ...780...21M}.
Irrespective of the cause of the CSM interaction, after the continuum emission has cooled down, the spectra of SN~2009ip  bear some resemblance to those of SNe~IIP 
and to those of SN~1998S \citep{fassia1998S}, the prototype of the sub-class of SNe~IIn that  SN~2009ip most closely resembles. 
The existence of SNe~Ibn \citep[e.g.][]{2007Natur.447..829P,2008MNRAS.389..113P} proves that  \textit{\textup{stripped}} core-collapse SNe can also explode within a CSM, although these objects have not been identified in an H-rich medium. 
Finally, thermonuclear supernovae have also been observed to strongly interact with a local CSM. The prototypical  object (hereafter SNe~Ia-CSM) was SN~2002ic \citep{2003Natur.424..651H}, followed by SN~2005gj \citep{2006ApJ...650..510A,2007arXiv0706.4088P}, SN~2008J
\citep{2012A&A...545L...7T}, and PTF11kx \citep{2012Sci...337..942D}. \cite{2013ApJS..207....3S} reported on the discovery of  additional SNe~Ia-CSM in archival and Palomar Transient Factory \citep[PTF;][]{2009PASP..121.1395L} data, raising the total number to 16 objects. 
Two more events have been reported since this study \citep{2014MNRAS.437L..51I,2014arXiv1408.6239F}.
Curiously, \cite{2012MNRAS.424.1372A} have shown that the distribution of SN~IIn locations with respect to star formation lies between those of SNe~Ia and those of SNe from more massive stars.

These observations 
suggest that SNe~IIn are not only characterized by a large observed heterogeneity, but also that this heterogeneity is due to profound physical differences.
Determining what kind of explosions/eruptions can hide under the CSM interaction, which masks them and labels them collectively as SNe~IIn,  is a fundamental question that is directly linked to our understanding of stellar evolution and mass loss. How many more SNe~IIn are, in reality, of thermonuclear nature? Can all SN~Ia sub-classes be linked with CSM interaction? What do the core-collapse explosions below H-rich CSM interaction look like? What hides below the faintest SNe~IIn?

These questions are also related to the search for the elusive progenitor systems of SNe~Ia:
in three out of four well-studied SNe~Ia-CSM (SNe~2002ic, 2005gj, and 2008J)
the underlying observed spectrum was best matched by SN~1991T, the prototype of a sub-class of luminous SNe~Ia  \citep{Filip91T}, while 
PTF11kx was spectroscopically similar to SN~1999aa, a SN intermediate to normal SNe~Ia and the 91T-like objects \citep{2004AJ....128..387G}.
Given that the number of 91T-like events among SNe~Ia is small, 9\% (18\%) in a volume (magnitude) limited survey \citep{2011MNRAS.412.1441L}, 
the question is raised whether this association is due to a physical reason (the progenitor system) or an observational bias.
The idea that SNe~Ia might originate from multiple channels \citep{mannucci2005} has gained support in recent years because of different lines of evidence \citep[e.g.][]{2006ApJ...648..868S,patat06Xuves,2011Sci...333..856S,2011Natur.480..348L,2012ApJ...752..101F,2012MNRAS.426.3282M,2013Sci...340..170W,2013ApJ...772...19F,2013MNRAS.436..222M}.
If real, the observed relation of 91T-like events to CSM interaction might indicate that these luminous explosions are related to single degenerate systems \citep{whelanIben_SD}.

Answering these questions is complicated because the SN spectra that are characteristic of the explosion types  and could be used to reveal their real nature are masked by the CSM interaction that washes out the differences.
\cite{2003Natur.424..651H} had to \textit{decompose} the observed spectrum of SN~2002ic into a 91T-like template superimposed on a smooth continuum to demonstrate the nature of the explosion. 
This led \cite{2006ApJ...653L.129B} to question the uniqueness of this solution arguing that a luminous SN~Ic could also provide a reasonable fit.
The procedure followed for SNe~2005gj and 2008J was similar, but in the case of PTF11kx no spectral decomposition was needed as the CSM interaction was not that strong. It is therefore obvious that for even stronger CSM interaction, the underlying SN spectrum will be even more diluted making it more difficult to identify and classify. 
This might be especially important for SNe that have been classified as Type~IIn based on a single spectrum or incomplete data.
The study of \cite{2013ApJS..207....3S} has proven that this is indeed the case and that the real fraction of SNe~Ia-CSM within SNe~IIn is certainly higher than the observed lower limit. 
Determining the real number requires a good knowledge of the conditions and parameters affecting the observed spectrum and our ability to classify it correctly.

In this paper, we try to address these questions by generating a large number of mock spectra of SNe interacting with a CSM.
This is done in a Monte Carlo (MC) way by varying a large number of parameters, including the SN type, brightness, and phase as well as the strength of the CSM interaction.
The generated spectra are distributed to different human classifiers and we try to quantify under what circumstances the true nature of the underlying SN can be recovered.
We parametrize our ability to classify a spectrum with respect to the flux ratio of the underlying SN over the CSM continuum.
By combining our results with observations, we place constraints on the nature of SNe~Ia that interact with a CSM, their fractions among SNe~IIn, and the parameter space where this occurs.
We extend our study to include different core-collapse SN types, even if it is not clear if such explosions (e.g. broad-lined Ic-CSM) exist in nature.

Our MC simulation is presented in Sect.~\ref{sec:sim}. The classification procedure and the classification results are discussed in Sect.~\ref{sec:clas}.
The results are analyzed separately for different kinds of SNe interacting with a CSM: Section~\ref{sec:Ia} presents our discussion on SNe~Ia, and 
Sect.~\ref{sec:Ic} and Sect.~\ref{sec:II} on stripped and H-rich core-collapse SNe, respectively. In Sect.~\ref{sec:faint} we speculate on what kind of explosion or eruption can be hidden below the faintest (e.g. $M > -17$) SNe~IIn. Section~\ref{sec:mod} discusses the physical motivation and the caveats of our simplified model. 
Our conclusions are summarized in Sect.~\ref{sec:conc}.

\section{Monte Carlo simulations}  
\label{sec:sim}

We have constructed spectra of SNe interacting with a CSM by adding the following components: (i) a SN template, (ii) a black-body (BB) continuum, (iii) narrow lines, such as those observed in SNe~IIn, (iv) extinction, and (v) noise. 
In Sect.~\ref{sec:mod}, it is shown that our simplified approach is a good first order approximation.
For each spectrum we vary a number of parameters by drawing them in a random way from distributions as described below.

(i) The SN templates used were those from \cite{2002PASP..114..803N} for normal, 91T-like and 91bg-like SNe~Ia as well as for Type Ibc SNe, broad-lined SNe~Ic (Ic-BL) and SNe~IIP.
The Nugent Ibc template contains He lines, similar to SNe~Ib. For this reason, we constructed an additional SN~Ic template based on the spectra of SN~2004aw \citep{2006MNRAS.371.1459T} and SN~2007gr \citep{2008ApJ...673L.155V}.\footnote{  throughout this paper the term Ibc will refer to the Nugent template, while Ic to new template without He.}
Each SN was assigned a different absolute maximum luminosity, assuming a dispersion around a central value for each SN type and drawing from a Gaussian distribution. The luminosity functions assumed can be seen in Table~\ref{tab:MC} together with other parameter distributions entering our simulation.
The average value for normal SNe~Ia is that from \cite{2010AJ....139..120F}. We have used a larger dispersion to also account for differences with other studies \citep[e.g.][]{2005ApJ...623.1011B}. It will be shown that our results are not as sensitive to absolute as to \textit{relative} quantities, such as the flux \textit{ratio} of the SN template to the BB continuum or the \textit{average difference} between normal SNe~Ia and 91T-like events. For the latter we have adopted 0.4~mag \citep[e.g.][]{2001ApJ...546..719L}.    
The reported absolute magnitude distributions of stripped SNe strongly depend on the assumed extinction and on their division in sub-classes  \citep{2002AJ....123..745R,2006AJ....131.2233R,2011ApJ...741...97D,2014arXiv1408.4084T}. Here, we have assumed a single luminosity function for all spectroscopically normal SNe~Ibc and Ic and a separate luminosity function for SNe~Ic-BL that are  brighter events \citep[e.g.][]{galama98}. The corresponding dispersions were set somewhat arbitrarily large in order to probe a wide region of the parameter space.
For Type~II SNe, represented here spectroscopically by the SN~IIP template\footnote{these terms will be used interchangeably throughout this paper.}, we have adopted a luminosity function based on 116 events \citep{2014ApJ...786...67A}.\footnote{since these simulations were made, \citeauthor{2014ApJ...786...67A} have slightly revised their luminosity function. The values used here were kindly provided to us prior to publication and are compatible with their final values.}
It was checked that convolved with a single extinction distribution (see below) the luminosity functions in Table~\ref{tab:MC} become compatible with the observed pseudo-luminosity functions of \cite{2011MNRAS.412.1441L}.

For each SN type, we used five different epochs corresponding to phases between $-$10 and $+$50 days past maximum (extending to $+$100 days for SNe~IIP) and
the spectra were scaled in flux with respect to their $V$-band maximum using the template $V$-band light curves provided with the Nugent templates. For the Ic template, we assumed the same light curve evolution as the Ibc template.
We did not modify the decline rate of the SN light curves, according to the maximum luminosity (i.e. no stretch correction was applied) as this precision is beyond the scope of our paper. In addition, we did not warp the template spectra to account for spectroscopic diversity  or different expansion velocities around the average values.

\begin{table}
\caption[]{Parameters of the MC simulation}
\label{tab:MC}
\centering
\begin{tabular}{llc}
\hline\hline
SN templates                   &                               &                                         \\
                                &       Ia (normal)                     &   $M_{V} = -19.12 \pm 0.30$             \\
                                &       91T-like                                &   $M_{V} = -19.52  \pm 0.30 $   \\
                                &       91bg-like                               &   $M_{V} = -17.92 \pm 0.50$             \\
                                &       Ibc \tablefootmark{a}           &   $M_{V} = -17.40 \pm 0.75$             \\
                                &       Ic \tablefootmark{a}             &   $M_{V} = -17.40 \pm 0.75$           \\
                                &       Ic-BL                                   &   $M_{V} = -19.14 \pm 0.50$             \\
                                &       II (IIP)                                &    $M_{V} = -17.02 \pm 0.99$            \\
\hline

Phases        &                               &                               \\
                                &   all Type~I          &   $-$10, 0, $+$10, $+$20, $+$50                            \\
                                &   IIP                 &   $-$5, 0, $+$25, $+$50, $+$100                           \\
                                &   probability                 &   1/8, 1/4, 1/4, 1/4, 1/8                              \\                              
\hline
                                
Black Body     &                               &                               \\
                                &      $T_{BB}$           &  6050$-$15750 $K $ \tablefootmark{b}          \\
                                &       $R_{BB}$        &   0.7$-$8.1 $\times$ 10$^{15}$ cm \tablefootmark{b}\\
\hline
                
Line Spectrum    &                                 &                                       \\
                                & line                          & H$_{\alpha}$, H$_{\beta}$ , H$_{\gamma}$,     \ion{He}{i} \tablefootmark{c} \\
                                & probability           &       1/1, 1/1, 1/2, 1/2                        \\
                                & FWHM broad            &  4000 $\pm$ 1000 km s$^{-1}$     \\
                                & FWHM narrow      &    400 $\pm$ 100   km s$^{-1}$                \\
\hline

Reddening              &                               &                                       \\
                                &       $A_V$           &  $\mu = 0$, $\sigma = 0.66$,        $A_V > 0$       \\
                                &       $R_V$           &  3.1                                                          \\

\hline          
                                
Noise        &                               &                                               \\
                                &       S/N     per pixel        &      5$-$60                         \\
 
\hline\hline
\end{tabular} \\
\tablefoot{
\tablefoottext{a}{Ibc refers to the Nugent template. Ic is the He-free template we constructed}.
\tablefoottext{b}{These distributions are phase-dependent. The limits given are those realized in the final simulation. See Fig.~\ref{fig:multipanel1}}.
\tablefoottext{c}{The \ion{He}{i} line is $\lambda$5876}.
}
\end{table}

\begin{figure*}
\begin{center}
\includegraphics[width=\textwidth]{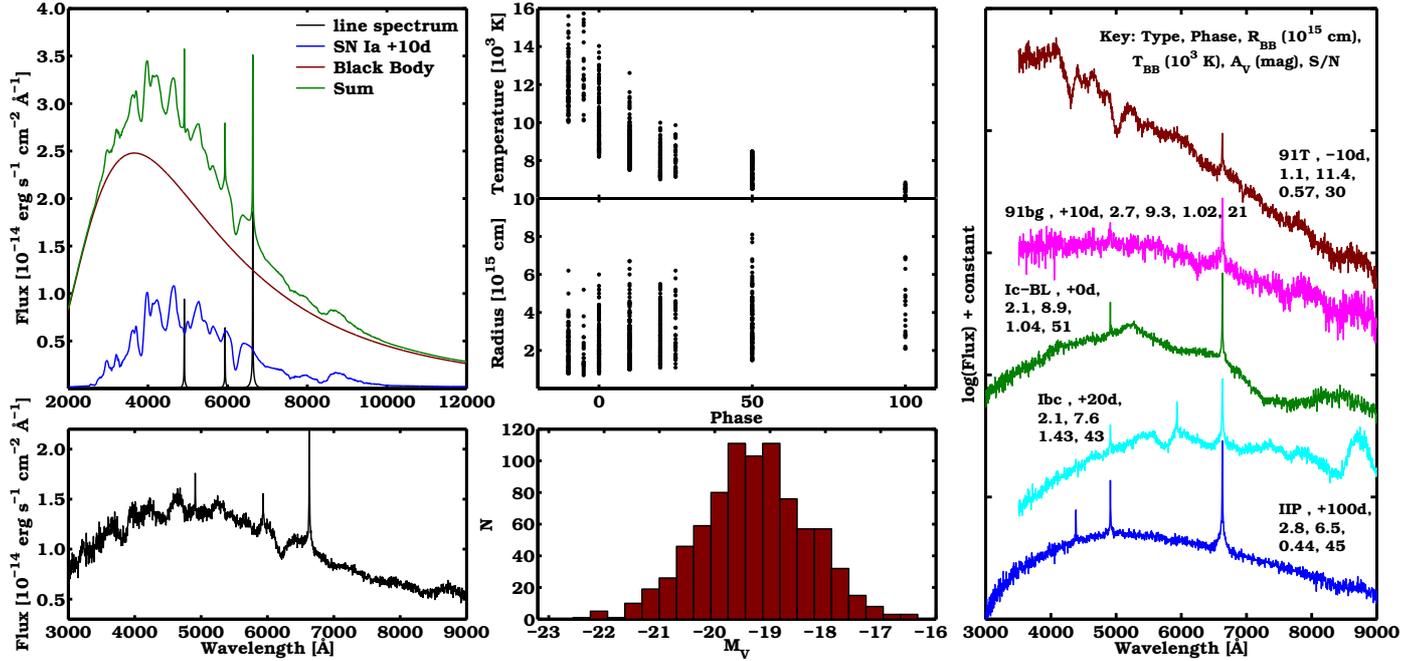}
\caption[]{\textbf{Left panels:} The top figure illustrates the procedure of spectral composition in our MC simulations for an example case. The composite spectrum is the sum of a SN template (here a SN~Ia at +10d), a BB continuum (here $R_{BB} = 3.2 \times 10^{15}$ cm and $T_{BB} = 8000$ K), and an emission line spectrum. In the bottom panel, we reddened the composite spectrum (by a total $A_V = 0.66$ mag) and added noise   (here S/N = 40) to obtain  the final spectrum in the (maximum) wavelength range 3000$-$9000 \AA. All fluxes are shown in the absolute scale.
\textbf{Middle panels:} The black-body radii and temperatures that were used in the MC simulation as a function of phase. The histogram in the bottom shows the resulting distribution of absolute magnitudes $M_V$. The obtained values range from  $-$22.5 to $-16$ and nicely overlap with the observed dispersion in SN~IIn luminosities.
\textbf{Right panel:} More example spectra taken from the final MC simulation and spanning a wide distribution of underlying SN types, phases, BB continuum strength, total extinction, and S/N. The values for these parameters are noted next to the spectra. 
The CSM contamination affects the spectral appearance.
The spectra are shown in a logarithmic scale and have been shifted by arbitrary constants for presentation purposes.
}
\label{fig:multipanel1}
\end{center}
\end{figure*}

(ii) The constraints on the BB temperature and radius represent the largest  (important) uncertainty that enters our simulation.
We have constructed a distribution based on values that have been typically reported in the SN~IIn literature \citep[e.g.][]{1993MNRAS.262..128T, fassia1998S, 2007ApJ...656..372G, 2008MNRAS.389..113P, Smith06tf,smith05ip,Smith06gy, Kiewe12,2012MNRAS.424..855K,2012ApJ...756..173S,2013A&A...555A..10T}.
The explosion time of SNe~IIn is often poorly constrained.
Nevertheless, it seems that temperatures around 8500~K are quite common, while values above 10000$-$12000~K and even exceeding 15000~K, have been observed, especially at early times (e.g. SN~1998S) or in superluminous events (e.g. SN~2006gy). 
At late times, SN~IIn temperatures do not seem to fall below 6000~K \citep{Smith06gy}.
We  assumed a distribution of temperatures around these values and an average cooling of the BB with time.
The temperatures for the 823 MC realizations used in this study are shown in Fig.~\ref{fig:multipanel1}.
Similarly, for the BB radius we assumed values of the order of 10$^{15}$~cm.
The distribution has been constructed so that the radii show a reasonable dispersion, an average increase with time and a possible turn-around at very late phases (Fig.~\ref{fig:multipanel1}).
It is very likely that the sample of well-studied SNe~IIn suffers from selection biases that are hard to quantify. 
It is therefore possible that our assumed distributions might not be adequate for studying ranges of the parameter space that have not been probed before, e.g. for very weak interactions or rare, possibly unrecognized, events.

(iii) The narrow lines have been constructed as the sum of two Lorentzians. Here we assumed a contribution from a broad and  narrow component. 
The first arises in the cold dense shell (CDS) that forms between the front and the reverse shock travelling in the CSM and is often referred to as the `intermediate' component.
The second is due to the unshocked CSM \citep[e.g.][]{1994ApJ...420..268C}.
Observationally, these components display a wide range in their observed widths (see e.g. references above).
Here we assumed a broad/intermediate component with FWHM of 4000 $\pm$ 1000~km~s$^{-1}$ and a narrow component with FWHM of 400 $\pm$ 100~km~s$^{-1}$ . 
The latter is also dependent on the spectral resolution of our spectrograph, which we assumed to be an average, low-resolution instrument. 
For simplicity, we only considered the strongest emission lines that are usually identified in SNe~IIn: H$\alpha$, H$\beta,$ and H$\gamma$.
We have also added He $\lambda$5876 in 50\% of the spectra. 
This was done independent of the SN type, although \cite{2013ApJS..207....3S} argued that this line rarely appears in SNe~Ia-CSM.

The simulation parameter controlling the strength of H$\alpha$ relative to the continuum was set so that H$\alpha$ is visually identifiable and spans different regimes from weak to strong. 
The equivalent width (EW) of H$\alpha$ was measured for the spectra used in this study and it was found to range between 23$-$51 \AA~(1 $\sigma$) with the most extreme EWs reaching 160 \AA.
This is consistent with the wide dispersion observed in SNe~IIn \citep[e.g.][]{2013A&A...555A..10T}.
For simplicity, the ratio of the Balmer lines was taken equal to Case B recombination \citep{1989agna.book.....O}. 
This ratio has been shown to deviate from this value for SNe~IIn, especially at later times \citep[e.g.][]{1993MNRAS.262..128T,2014AJ....147...23L,2013ApJS..207....3S} but this approximation is adequate for our purposes. 
No shifts with respect to the central wavelength of the lines were assumed.

(iv)  The SN template, the BB continuum, and the narrow-line spectrum were added together to obtain a composite spectrum. Subsequently, the composite spectrum was reddened assuming a half-Gaussian distribution for $A_V$ with $\mu = 0$ and $\sigma = 0.66$ mag. We decided to use the same average extinction towards different kinds of SNe, although it  is possible that this is not the case in nature. For instance, it has been argued that 91T-like objects might suffer from higher extinction than normal SNe~Ia \citep[e.g.][]{2001ApJ...546..719L,2001ApJ...546..734L}, while it is obvious that for SNe~Ibc (and sub-classes) the assumed extinction plays an important role in the derived luminosity functions \citep{2006AJ....131.2233R,2011ApJ...741...97D}.
The ratio of total to selective extinction $R_V$ was kept equal to  3.1, although there are both theoretical arguments \citep{2008ApJ...686L.103G,2011ApJ...735...20A} and observational evidence \citep{2010AJ....139..120F,2012A&A...545L...7T} suggesting that this value is different (lower) for SNe~Ia interacting with a CSM.

By convolving the assumed luminosity functions  for the different SN types (Table~\ref{tab:MC}) with this extinction distribution, the results become compatible with the observed pseudo-luminosity functions of \cite{2011MNRAS.412.1441L}.
This indicates that our assumptions are realistic to a first order.
In addition, it will be shown that extinction does not play any significant role in the correct classification of our spectra.

(v) Finally, we added Gaussian noise to the spectra with S/N ranging from 5 to 60 per pixel.
To simulate more realistic conditions, the noise was increased in the bluer ($<$~4500~\AA) and redder ($>$~7500~\AA) parts of the spectra. 

In general, we have tried to simulate both realistic physical properties and observing conditions. 
We generated more spectra around maximum and post-maximum phases, because there are relatively more SN discoveries during these phases than at pre-maximum or very-late phases. 
According to what is usually reported in classification circulars, the spectra were cut in the blue part at 3000, 3500, or 4000 \AA, with equal probability. 
The red cutoff was taken at 9000 \AA,\ but this wavelength range is not crucial for our project (as long as H$\alpha$ is included).
Our purpose was to  obtain reasonable statistics for all SN types rather than to follow the observed SN rate \cite[e.g.][]{2011MNRAS.412.1441L}.
As a check, a small number of generated spectra did not include any underlying SN template, but only the other components.
All spectra were redshifted to $z=0.01,$ but this was just an arbitrary choice made for simplicity. 
The cosmology used was $H_0 = 71$ km s$^{-1}$ Mpc$^{-1}$, $\Omega_{\rm{m}} = 0.27,$ and  $\Omega_{\Lambda} = 0.73,$ but we stress that our conclusions are independent of the distance. 

The parameters of the simulation are summarized in Table~\ref{tab:MC}.
The simulation procedure is illustrated in Fig.~\ref{fig:multipanel1} together with some example spectra.

\section{Classification of the simulated spectra} 
\label{sec:clas}

\begin{figure*}
\begin{center}
\includegraphics[width=\textwidth]{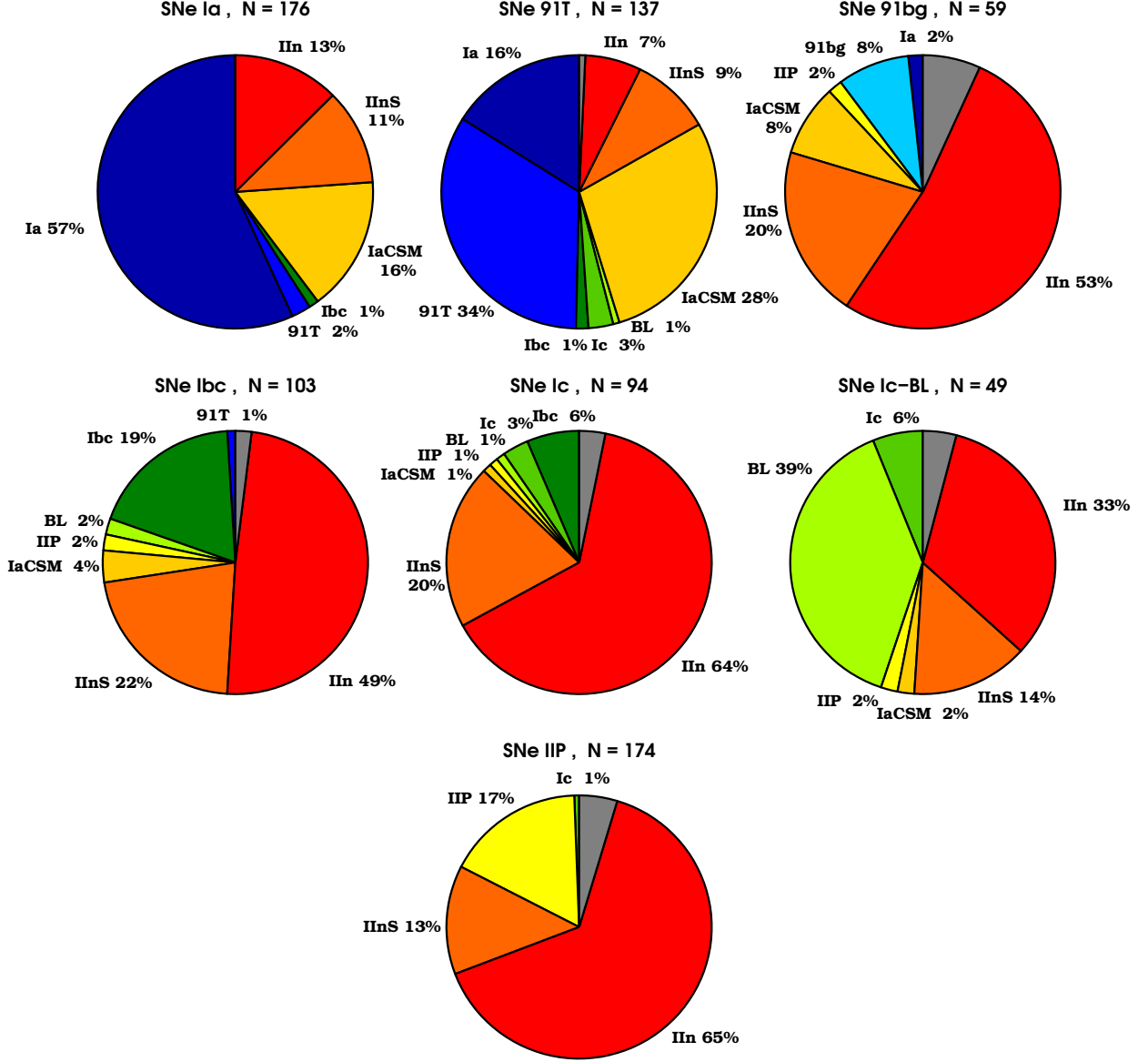}
\caption[]{Pie charts for each spectral type that was used as input to the simulation (Ia-normal, 91T, 91bg, Ibc, Ic, Ic-BL, and IIP) and showing how these were typed by the human classifiers. The total number of simulated spectra per spectral type is indicated above the pie charts. All percentages have been rounded to the closest integer.  In addition to the seven input types, the following classifications were allowed: Ia-CSM, IInS, and IIn (see Sect.~\ref{sec:clas}). The grey colour corresponds to a few SNe that were classified as `none'. The recovery rate is higher  among bright SNe. Fainter sub-types are more often classified as SNe~IIn (including IInS). See text for more details.
}
\label{fig:inputpie}
\end{center}
\end{figure*}

\begin{figure*}
\begin{center}
\includegraphics[width=\textwidth]{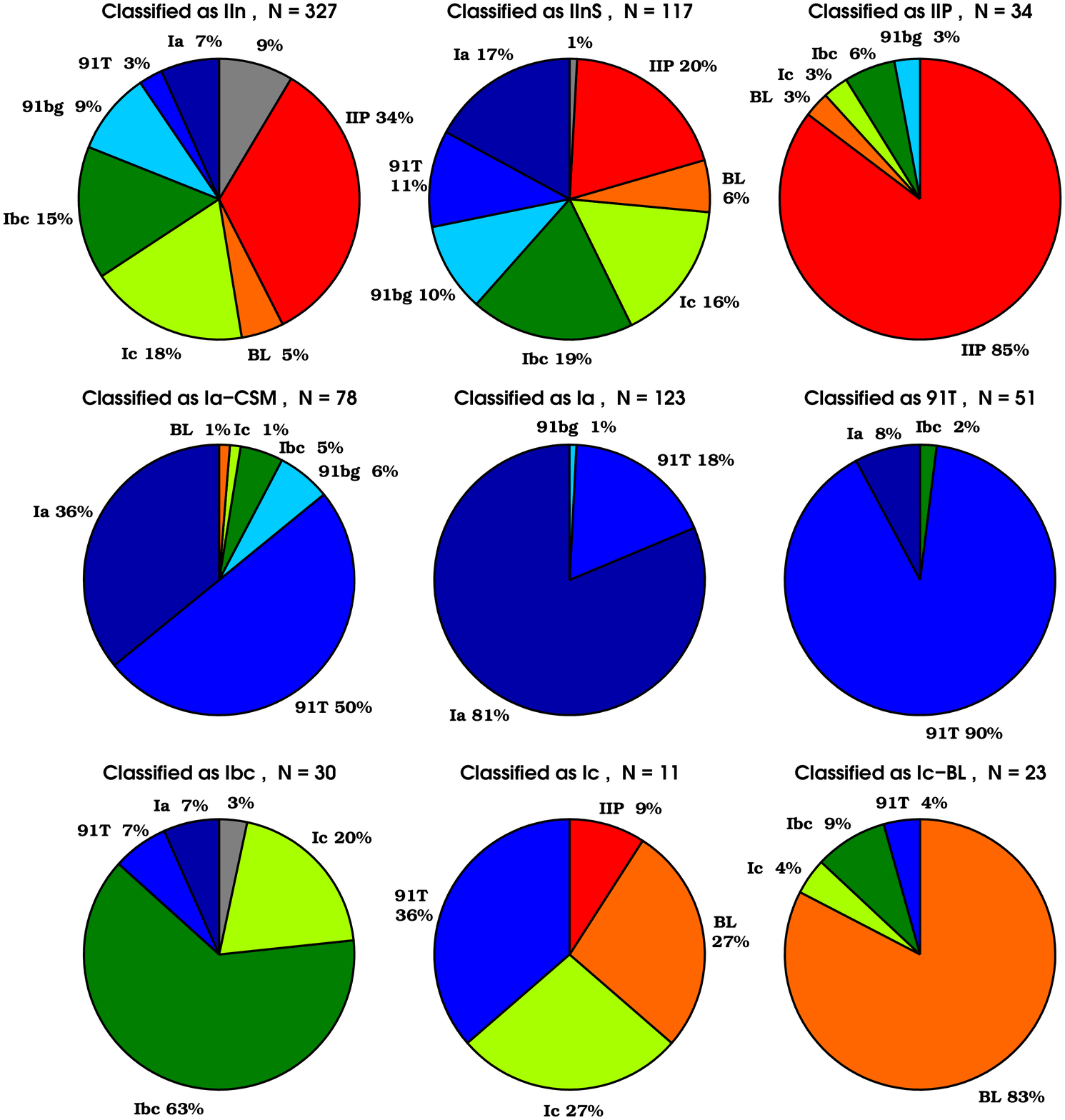}
\caption[]{Pie charts showing the real nature of objects that were given a certain classification.
The number of spectra is indicated above the pie charts. All percentages have been rounded to the closest integer.
The first row shows the distribution of the input SN templates for spectra that were classified as  IIn, IInS, and IIP, respectively. 
The second row shows the same for all spectra that were typed as thermonuclear. 
No pie chart is shown for the events classified as 91bg-like. Five simulated events were classified as such 
and the underlying SN was indeed 91bg-like in all cases. 
The third row shows the real nature of the SNe that were typed as various stripped core-collapse SNe.
The contamination from core-collapse SNe in the spectra classified as thermonuclear  reached  7\% in the Ia-CSM pie, with one more stripped event classified as 91T-like.
Up to 14\% of the objects classified as stripped core-collapse explosions were really thermonuclear, with the contamination being lower for Ic-BL and higher for SNe~Ic. 
The grey colour indicates simulated spectra where no SN template was included.
As expected, most of these were classified as IIn.
}
\label{fig:outputpie}
\end{center}
\end{figure*}

We have constructed 823 spectra with the method described in Sect.~\ref{sec:sim}.
These spectra were distributed among ten authors of this paper (EYH, JJ, GL, JN, TP, JMS, MDS, JS, FT, DX) for classification. 
The purpose of this experiment was to investigate in which regions of the parameter space it was possible to classify an object according to its true nature.
The classifiers were given the following instructions: they could classify a supernova as a 
Ia (normal), 91T-like, 91bg-like, Ibc, Ic, Ic-BL, or IIP when they believed that this was the accurate type.
In addition, the classifiers were given the following options: they could classify a SN as Ia-CSM if they thought that the underlying spectrum was of thermonuclear nature but it was not possible to distinguish the exact sub-type. All the other SNe should be classified as IIn. 
However, an additional class was introduced for the needs of this project: the classifiers were asked to use a \textit{IInS} designation (where the `S' stands for `structure') when the spectrum demonstrated some evident structure (or wiggles) besides the continuum and emission lines. 
This was done to characterize an intermediate regime where something \textit{\textup{just}} starts appearing over the continuum, but the information is not enough to warrant any classification different than IIn. 

 Apart from that, the classifiers were given full freedom  to use any tool that they found convenient, such as SNID \citep{2007ApJ...666.1024B}, GELATO \citep{2008A&A...488..383H}, or Superfit \citep{2005ApJ...634.1190H}, as well as their personal criteria. In general, the classifiers were asked to act as if they had to classify these spectra in reality and submit a circular. They were given the option to provide brief comments for each SN for anything they would consider worth noting. Individual comments were used to improve the procedure and facilitate the analysis, but they will not be discussed      in this paper, with a few exceptions. If the classifiers thought that a spectrum did not look realistic, they were asked to provide a `none' classification. 
 The advantage of having many classifiers, rather than one, is evident: 
 this way it was secured that the results do not depend on a single individual (small differences were found between classifiers), but that the human factor (including experience and personality) and a wider range of external conditions (such as tiredness or time constraints) would enter the procedure. 
Both expert spectroscopists, regular SN classifiers and novice students (after receiving training)  participated in this experiment to make it more realistic.
We note that the classifiers did not have access to the brightness of the targets but only provided their input based on the spectral shape.

In the rest of this section, we discuss some general results from the classification procedure. 
We provide statistics on the general  `correct versus wrong' classification, but this is only done with respect to the SN type and 
independent of the other simulation parameters (S/N, extinction, etc).
A more detailed analysis is given in the following sections.

Figure~\ref{fig:inputpie} contains pie charts for each of the seven input spectral types  
showing how these were classified. We observe that the majority of normal SNe~Ia ($>$ 50\%) were classified correctly.
A lower percentage of events were classified as Ia-CSM (16\%), IInS (11\%) or IIn (13\%) due to the increasing contribution of the continuum. There were only two real misclassifications\footnote{By \textit{real} misclassification we mean the assignment of an incorrect SN type. This is not the case for SNe~IInS and IIn, as many times this is the best possible classification.} (Ibc) in a total of 176 events.

The `success rate' of correct typing among 91T-like objects is lower (35\%), but this is due to a `leak' towards Ia (16\%).
The reason behind this is time evolution: the spectra become very similar after $\sim$15 days past maximum \citep[e.g.][]{1997ARA&A..35..309F,2001ApJ...546..734L}.
Few 91T-like objects were actually classified as 91T-like at phase +20, and none later than that.
In contrast, all 91T-like objects that were classified as Ia were at phases +10 or later.
The fractions of 91T-like events that were classified as  Ia-CSM, IInS,  and IIn  were 29\%, 10\%, and 7\%, respectively. 
The larger ratio of Ia-CSM to IIn classifications, compared to normal SNe~Ia, is because 91T-like objects are on average brighter than normal events. 
Classifiers falsely attributed  3\% of 91T-like objects to stripped core-collapse SNe.

The lower luminosity of 91bg-like objects leads to a much lower recovery ratio: only 8\% in our simulation were identified with their correct sub-type, while almost 75\% would have been classified as SNe~IIn (including IInS). 
It is interesting to point out, however, that for 3/6 events that were classified as Ia or Ia-CSM, there were associated comments by the classifiers hinting at some similarity with SN~1991bg.

The SNe~Ic-BL also score relatively high in correct identifications ($\sim$40\%), which is probably because of their high luminosities (comparable to SNe~Ia) and their characteristic broad features. It is not clear if SNe~Ic-BL can explode within a dense H-rich CSM, but the fact that they are likely easy to identify places strict constraints on their frequency of occurrence  in nature. 
The fainter core-collapse SNe (Ibc and IIP) score lower percentages and it is thus more likely that such events constitute a fair fraction of SNe~IIn.
Two things are worth mentioning: first, there is only one real misclassification in 174 IIP input spectra (all other objects were either classified as IIP, IInS, or IIn). Second, the number of SNe~Ibc and Ic that were simulated to interact with a CSM and were given a thermonuclear classification (Ia-CSM or 91T-like) is small: 5\% among Ibc and 1\% among Ic in a total of almost 200 trials. This suggests that despite the spectral similarities between SNe~Ia and Ibc \citep[e.g.][]{2006ApJ...653L.129B}, it is usually possible to distinguish between the two. Thus, the objects classified as SNe~Ia-CSM \citep[e.g.][]{2003Natur.424..651H} are unlikely to be misclassified SNe~Ibc.

Figure~\ref{fig:outputpie} shows the real nature of the objects that were given a certain classification. 
The objects classified as Ia-CSM were indeed mostly of thermonuclear nature and the contamination from stripped core-collapse SNe is small (7\%). 
All objects classified as 91T- and 91bg-like indeed belonged to these classes, except in one case (accuracy $>$ 98\%). This indicates that classifiers were always confident when assigning a
sub-type to SNe~Ia. Normal events demonstrate a (small) contamination (19\%), but this was always from other thermonuclear sub-types and it is probably because of a conservative approach by the classifiers and because of the difficulty in typing 91T-like events at phases past +20 days.

The errors among objects that were classified as stripped core-collapse SNe were larger. Almost 20\% of all objects that were given an Ibc or Ic designation were, in reality, of thermonuclear nature. The contamination is bigger among SNe that were characterized as (He-free) SNe~Ic, 
and in the vast majority of the cases it is caused by 91T-like events rather than normal SNe~Ia. This is expected as it is these sub-types that demonstrate the biggest spectroscopic similarities (see Section~\ref{sec:Ic}). 
We note that the classifiers often preferred a more conservative Ibc classification rather than Ic, even if the SN did not show He lines.
Broad-lined events show a much smaller contamination (4\%), probably because of their more distinguished spectral shape.

An interesting comment by classifiers is that, in many cases, the CSM narrow lines would have simply been considered host galaxy lines. 
We do not examine host galaxy contamination here. Based on the results presented in the following sections, however, we have good reason to believe that many of the conclusions concerning the correct SN typing would apply in the case of host contamination as well.

\section{Thermonuclear SNe}    
\label{sec:Ia}

After presenting the fraction of successful classifications per SN type interacting strongly with a CSM in our simulation, we now concentrate on studying the reasons behind these classifications and their implications.  In this section we focus on thermonuclear SNe.
We note that this paper only deals with \textit{\textup{strong}} CSM interaction (SNe~Ia-CSM).
SNe~Ia have also shown signs of CSM interaction in a weaker regime \citep[e.g.][]{patat06Xuves}, but this is not studied here.

\subsection{No luminosity bias in favour of 91T-like events and against normal events}
\label{sec:nobias}

\begin{figure}
\begin{center}
\includegraphics[width=\columnwidth]{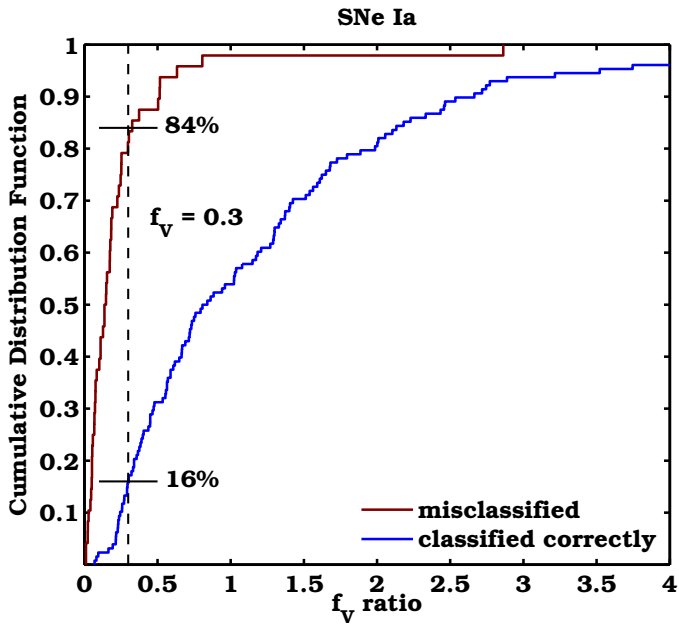}
\caption[]{Distributions of simulated SNe~Ia (normal) interacting with a CSM with respect to the flux ratio $f_V$. The objects classified correctly have larger $f_V$ ratios and the two distributions are statistically different at a high significance ($p \sim 10^{-15}$ in a KS test).
All spectra that were given a Ia-CSM designation have been considered to be correctly classified.
The dashed line indicates that less than 16\% of the simulated events were correctly (wrongly) classified at flux ratios below (above) $f_V = 0.3$.
}
\label{fig:Iacorwrong}
\end{center}
\end{figure}

There are conclusions to be drawn already based on the results presented in Sect.~\ref{sec:clas}. First, it was shown that it is not particularly difficult to identify bright thermonuclear SNe (Ia-normal and 91T-like) interacting with a CSM. More than 75\% of the events were either identified correctly or as Ia-CSM, while only 16-24\% were typed as IIn (including IInS). 
Second, we do not see any bias in preferentially correctly classifying   91T-like objects versus normal Ia due to their higher luminosities (rather the opposite due to time evolution). 
It is thus demonstrated that the observed association between Ia-CSM objects and 91T-like events \citep{2003Natur.424..651H,2006ApJ...650..510A,2007arXiv0706.4088P,2012A&A...545L...7T}, is not due to a luminosity bias.
For the extended sample of \cite{2013ApJS..207....3S} , it is not straightforward to confirm (or reject) this association as many of their objects are at lower S/N or at phases sufficiently late to make a distinction.
Nevertheless, \cite{2013ApJS..207....3S} note a preference for objects with early spectra to resemble the luminous SN~1999aa (see also Fig.~\ref{fig:real_data_comp}).
A chance coincidence is unlikely because of the small fraction of 91T-like objects among SNe~Ia \citep[9-18\%;][]{2011MNRAS.412.1441L}.
Hence, this observed association must be due to a  \textit{\textup{real physical reason}}, linked to the progenitor system.  
We therefore propose that the sub-class of 91T-like SNe~Ia come from the single degenerate channel, as it is more likely for this channel to generate a dense CSM \citep[e.g.][]{2012ApJ...761..182M}.

This could be in line with the well-known preference of the brightest SNe~Ia to occur in late-type, high star-forming galaxies  
\citep[e.g.][]{2000AJ....120.1479H}.
The hosts of SNe~Ia-CSM were all found to be either late-type spirals or dwarf irregulars, implying a young stellar population \citep{2013ApJS..207....3S}. 
Our suggestion does  not imply  that all 91T-like SNe should show signs of CSM interaction (and in fact they do not; e.g. \citealt{2014MNRAS.445...30S}), as this depends on the time of explosion relative to the mass-loss history of the companion star.
The statistical study of \cite{2013MNRAS.436..222M} found that SNe~Ia with blue-shifted \ion{Na}{i}~D absorption (and thereby evidence for local  \textit{\textup{weak}} CSM interaction) also show a preference for late-type galaxies.

\subsection{The dependence on the ratio of fluxes contributed by the SN and the continuum}
\label{sec:fv}

To quantify our discussion, we introduce the following parametrization:

\begin{equation}
f_X = \frac{F_{SN}}{F_{BB}}
\label{eq:fx}
,\end{equation}

where $F_{SN}$ and $F_{BB}$ are the fluxes from the SN and the continuum, respectively, and $X$ denotes the bandpass these fluxes are measured in, e.g. $f_V$ is the flux ratio in the $V$ band.
The advantage of using the ratio of fluxes is that our conclusions become independent of our assumptions on the absolute quantities 
entering our simulation (e.g. $M_V$, $T_{BB}$, $R_{BB}$) and can be generalized. This quantity is trivial to calculate for each spectrum in our simulation and we recorded the value of this ratio in the $B$, $V$ and $R$-bands.

Figure~\ref{fig:Iacorwrong} shows the distributions of normal SNe~Ia that were correctly and wrongly classified with respect to the ratio $f_V$. 
Here we have included all events that were given an Ia or Ia-CSM classification in the `correct' sample, that comprises  128 objects.
The wrongly classified sample includes IIn, IInS objects, and real misclassifications ($N =$ 48). 
It is not surprising to see that the successfully classified SNe have larger $f_V$ ratios: when the flux ratio is large, the spectral features of the underlying SN are less diluted by the continuum and this makes them more easy to identify and classify.
The two distributions are clearly drawn from a different parent distribution: a Kolmogorov-Smirnov (KS) test gives  $p \sim 10^{-15}$.
Below, we will show that there is no other parameter in our simulation that significantly affects the success of a classification and, therefore, we will suggest that the ratio $f_V$ is the \textit{\textup{sole important parameter}} determining whether the nature of a SN interacting strongly with a CSM will be accurately recovered. 

Furthermore, there seems to be a ratio threshold that is directly linked with the probability of recovering the nature of the underlying SN and this seems to lie around $f_V \sim 0.3$ for SNe~Ia. In Fig.~\ref{fig:Iacorwrong} it is shown that  less than 16\% of the objects that were correctly classified have $f_V < 0.3$, while less than 16\% of the objects that were wrongly classified have $f_V > 0.3$. Similar limits exist for 91T-like and 91bg-like objects: only 15/107 and 4/11 of the correctly classified objects were at  $f_V < 0.3$. Since these values are consistent with what we derived for normal SNe~Ia, with poorer statistics, we conclude that $\sim$0.3 is a representative threshold for the ratio $f_V$.

\begin{figure}
\begin{center}
\includegraphics[width=\columnwidth]{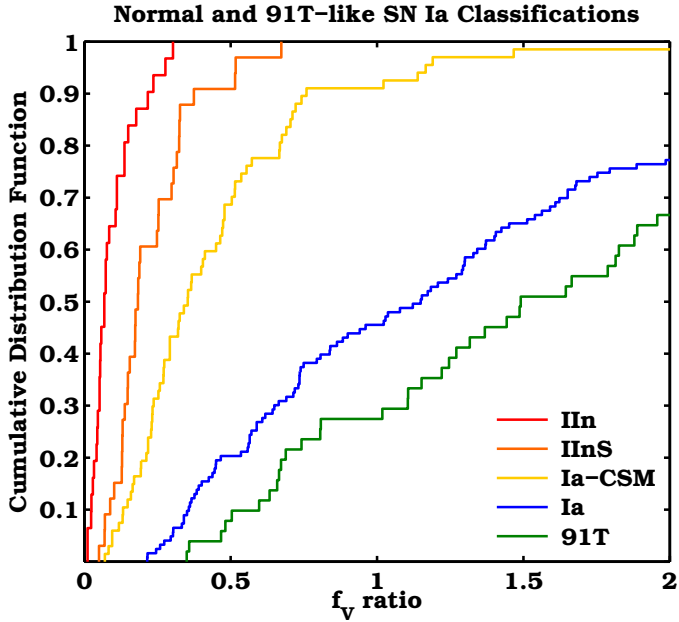}
\caption[]{Distributions of all the simulated normal and 91T-like SNe~Ia interacting with a CSM and how these were classified depending on the flux ratio $f_V$. As $f_V$ decreases (the contribution of the BB continuum increases), the SN features become more diluted and the SNe are progressively classified as Ia-CSM, IInS, and IIn.
All distributions are mutually statistically incompatible. 
The difference between the Ia and the 91T distributions is explained by the fact that 91T-like events are brighter, on average, than normal SNe~Ia (by 0.4 mag in this simulation).
The figure zooms in $f_V \leq 2$.}
\label{fig:IIn91Tseries_fv}
\end{center}
\end{figure}

\begin{figure}
\begin{center}
\includegraphics[width=\columnwidth]{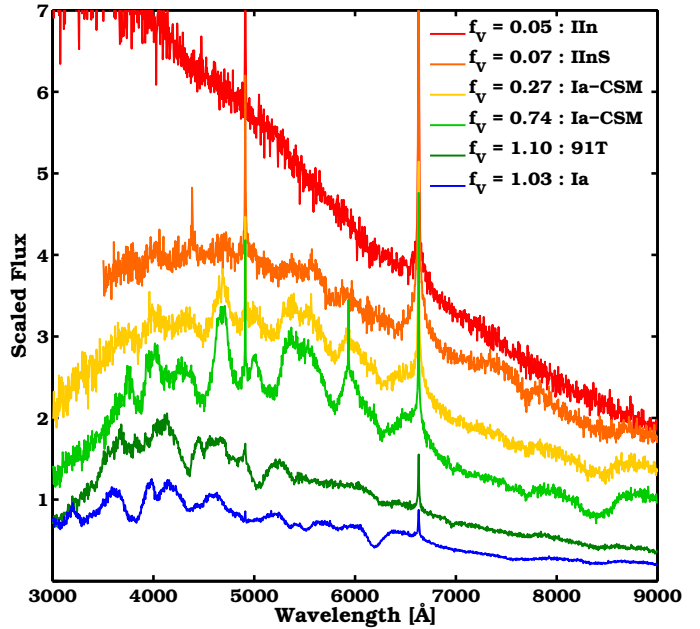}
\caption[]{Representative spectra from the MC simulation, visualizing the effect of $f_V$ in the final spectrum appearance and classification.
A normal SN~Ia and a 91T-like event are shown at maximum light with $f_V \sim 1$. Even at this 1:1 level, the CSM contamination is barely visible in the spectra. 
From objects that were classified as Ia-CSM we show two  cases with $f_V = 0.74$ and $0.27$.
At lower flux ratios, it is not possible to distinguish the spectral features  and the SNe were classified progressively as IInS and IIn.
The spectra have been scaled by arbitrary constants for presentation purposes. 
}
\label{fig:exampleseries}
\end{center}
\end{figure}

The importance of the flux ratio $f_V$ is further illustrated in Fig.~\ref{fig:IIn91Tseries_fv} that shows its progressive effect on objects classified as IIn, IInS, Ia-CSM, Ia-normal, and 91T. To allow for a fair comparison, only objects that initially contained an underlying normal Ia or 91T-like event have been included (e.g. simulated SNe~IIP that were classified as IIn have been excluded from this plot). As the flux ratio decreases, it becomes more and more difficult to correctly classify the SN. First, we obtain an object that looks like a Ia-CSM, then a IIn with some structure/wiggles (IInS) and finally a IIn. 
SNe classified as IIn occupy the region  $0.03 < f_V < 0.15$ (excluding 16\% outliers from both sides), IInS are found between 0.10 - 0.32, Ia-CSM between  0.17 - 0.69, Ia-normal between  0.42 - 2.33, and 91T-like between 0.66 - 3.00.
None of these distributions is statistically compatible with the others. 
The highest p-value in a KS test is found between the Ia-CSM and IInS distributions ($ 2\times 10^{-4}$). 

Representative examples of spectra corresponding to these classes are shown in Fig.~\ref{fig:exampleseries}. The objects classified as Ia-normal and 91T-like have been selected to both have $f_V \sim 1$ and a phase at maximum light to highlight the well-known spectroscopic differences between these two classes. Even at this level of contamination, with equal contribution from the BB continuum and the SN, one can hardly notice the CSM interaction. The spectra look normal, although slightly over-luminous if the classifier had access to accurate photometry, and the narrow lines could easily have been attributed to host galaxy contamination.
From objects that were classified as Ia-CSM, we show two limiting cases (both 91T-like at +20d) with $f_V = 0.74$ and $0.27$. In the first simulated spectrum, there is also 
 \ion{He}{i} $\lambda$5876 emission, although \cite{2013ApJS..207....3S} have argued that this is rare in SNe~Ia-CSM.
Finally, the spectrum of the object classified as IInS 
presents some wiggles, but the features are  diluted by the continuum  that it is impossible to uncover its thermonuclear nature.
In reality, this object would have been classified as a SN~IIn, although the differences between this object and the IIn with lower $f_V$ are evident.

 \begin{figure*}
\begin{center}
\includegraphics[width=\textwidth]{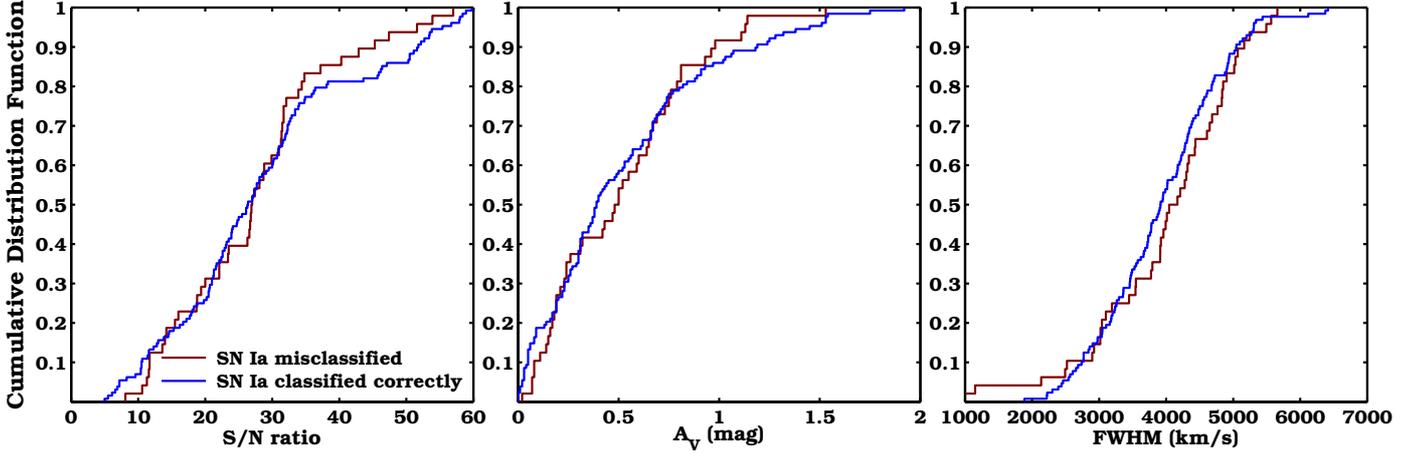}
\caption[]{Same as in Fig.~\ref{fig:Iacorwrong}, for the S/N ratio (left), total extinction $A_V$ (middle) and the width of the broad/intermediate component in the emission lines (right). 
The KS test p-values are 0.89, 0.59 and 0.49 respectively.
}
\label{fig:Ia_SNRAvFWHM}
\end{center}
\end{figure*}

\begin{figure*}
\begin{center}
\includegraphics[width=\textwidth]{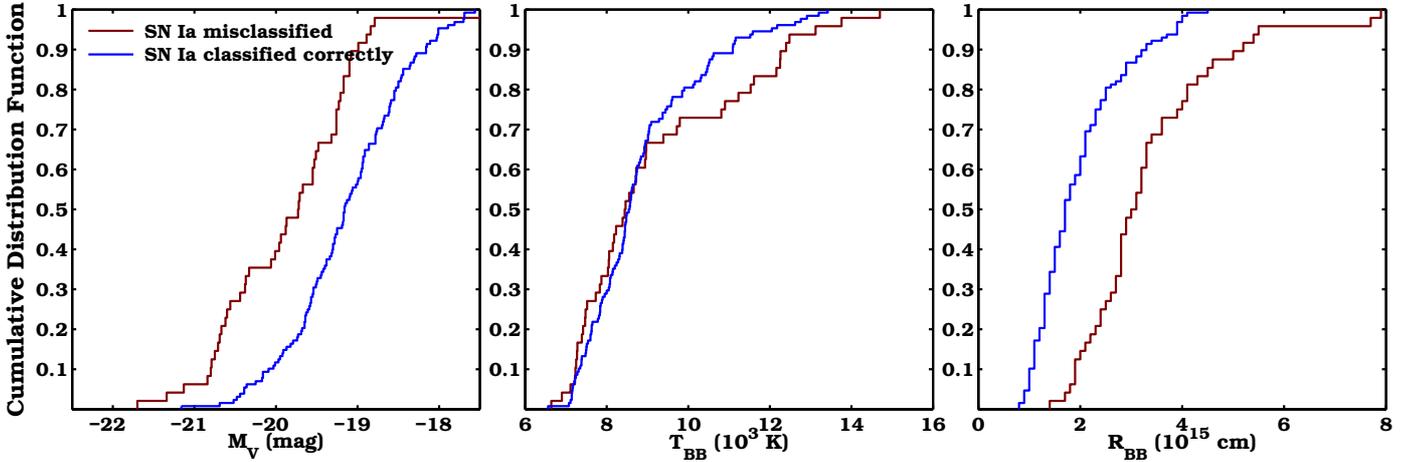}
\caption[]{Same as in Fig.~\ref{fig:Iacorwrong}, for the total $M_V$ of the final spectrum (left), the BB temperature (middle), and the BB radius (right).
All these parameters are degenerate with $f_V$ that has the advantage of being independent of absolute quantities. 
As expected, the objects that are harder to classify are the ones that reach higher total luminosities, since they have a higher BB contribution.
It is shown that the BB radius is more critical for the correct identification of the underlying event than the BB temperature, although a difference exists at T $>$ 8500 K.
}
\label{fig:Ia_MvBB}
\end{center}
\end{figure*}

Simulation parameters that are not degenerate with $f_V$ do not affect the classifications significantly. 
This is illustrated in Fig.~\ref{fig:Ia_SNRAvFWHM}, which shows the distributions of correctly and wrongly classified SNe~Ia with respect to the S/N ratio, the total extinction $A_V$, and the width of the broad/intermediate emission lines. 
Both visual inspection and KS tests reveal that the corresponding distributions are indistinguishable, indicating that none of these parameters play any role in the correct identification of the spectrum. This is at least true within the parameter range covered by our simulation (5 $<$ S/N $<60$, $0 < A_V <2$ mag, $1000 <$ FWHM $<7000$ km~s$^{-1}$). 
A possible caveat is that the simulated extinction was a total extinction (Galactic and host galaxy component) that is applied equally to the SN and the continuum. If, however, the SN and the CSM experience different reddening (or extinction laws), this might affect their flux ratios and thereby the appearance of the spectrum.  
This scenario is not unthinkable \cite[e.g.][]{2008ApJ...686L.103G}, but it would only become important for very highly reddened events and in the bluer bands. 
Since we see no evidence at all for any dependence from a total $A_V$ and no significant difference in whether we use $f_B$ rather than $f_V$, we expect any effects from a `broken' $A_V$ to be minimal.
Similarly, a successful  classification does not depend on the width of the narrow 
component, the existence of  \ion{He}{i}, or on the parameter controlling the strength of the emission lines over the continuum (and therefore their EW).
Although potentially very interesting, a study with respect to the SN phase is not possible because this is strongly dependent on the assumptions that we made on the BB evolution with time.
 
Although the absolute luminosity of the SN and the BB temperature and radius are largely degenerate with $f_V$, valuable conclusions can also be drawn by studying the dependence on these parameters. This is done in Fig.~\ref{fig:Ia_MvBB}  for normal SNe~Ia. As expected, the SNe that were misclassified are brighter than those that were classified correctly ($p = 1.5 \times 10^{-4}$) because of the increased contribution by the BB continuum diluting the spectral features. It is stressed that the $M_V$ plotted is the \textit{total} $M_V$ (and not that of the SN template) and that it phase-dependent, i.e. it does not refer to maximum light. 

Breaking the BB contribution into its fundamental parameters and studying them separately, we come to the following conclusion:
the distinguishing difference does not seem to be due to the temperature ($p = 0.30$) but principally  to  the radius ($p \sim 10^{-10}$). 
Only 20\% of the SNe that were identified correctly had $R_{BB} > 2.4 \times 10^{15}$~cm, while more than 80\% of the misclassified SNe are in the same radius range. 
The temperature effect becomes important only at the highest temperatures considered here: above 8500 K, the correctly classified spectra are systematically found at lower temperatures than those that were misclassified. We note that this is approximately the temperature where the peak of the BB function starts moving outside our average wavelength range ($\sim$ 3400 \AA). 
The difference is above the 2$\sigma$ level ($p = 0.02$) only if the temperatures above 8500 K are considered.

\subsection{Magnitude range of Ia-CSM supernovae}

Based on our findings we are able to place constraints on the magnitude range where one expects to find Ia-CSM supernovae.

If we assume that the flux from the narrow emission lines is negligible, the total flux received is $F_{tot} = F_{SN} + F_{BB}$. Using the definition of $f_X$ in equation~\ref{eq:fx}, it is easy to show that:

\begin{equation}
M_{tot} = M_{SN} - 2.5\log{(1+\frac{1}{f_X})}
\end{equation}

where $M_{tot}$ is the absolute magnitude of a SN interacting with a CSM and $M_{SN}$ is the absolute magnitude of the supernova component alone (i.e. without the continuum contribution). Both $M_{tot}$ and $M_{SN}$ are measured in the same bandpass $X$.

\begin{figure}
\begin{center}
\includegraphics[width=\columnwidth]{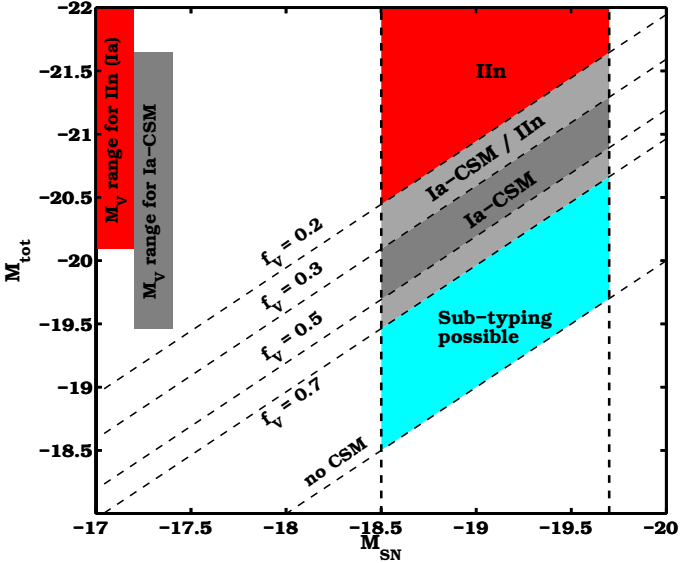}
\caption[]{Graph showing the total maximum luminosity of a SN interacting with a CSM (i.e together with the continuum contribution), as a function of  the underlying SN maximum luminosity and the ratio $f_V$. The graph focuses on the magnitude region occupied by SNe~Ia (vertical dashed lines). As shown before (e.g. Fig~\ref{fig:IIn91Tseries_fv}), the spectrum of the SN interacting strongly with the CSM will receive different classifications,  depending on the value of $f_V$ (diagonal dashed lines). SNe in the red-colored area will be classified as a IIn (including IInS in this project) and SNe in the dark-grey area as Ia-CSM. Objects in the cyan area can be classified with their exact sub-type (e.g. 91T-like), while the light-grey areas represent transitional zones where more classifications are possible.
It is therefore possible to determine the (un-extincted) magnitude region for SNe~Ia-CSM ($-19.5 > M_V > - 21.6$) and the lower luminosity limit for a SN~IIn that can hide a SN~Ia ($M_V < -20.1$). These regions are illustrated by the red and grey vertical bars on the top left. 
}
\label{fig:IaIInAllowed}
\end{center}
\end{figure}

Figure~\ref{fig:IaIInAllowed} shows a graph relating these quantities, where the lines of different $f_V$ appear as diagonals.
In the previous section, we derived that a correct identification is usually possible at $f_V \gtrsim$ 0.3, but that Ia-CSM classifications are possible all the way down to 0.2. At higher flux ratios, objects were given Ia-CSM classifications up to 0.7, although accurate sub-typing starts being possible already at  $f_V \sim$ 0.5. These limits divide the parameter space of Fig.~\ref{fig:IaIInAllowed} in 5 different regions (2 of which are transitional) reflecting the appearance of the final spectrum.

Assuming a luminosity function for $M_{SN}$ it is therefore possible to place constraints on the $M_{tot}$ of Ia-CSM SNe.
The power of working with the flux ratio is precisely that this analysis can be generalized for any luminosity function and remains independent of the luminosity function assumed in our simulation.
To allow for a better comparison with the observational work of \cite{2013ApJS..207....3S}, we have assumed the same range for SNe~Ia, which obeys the magnitude-stretch relation \citep{1993ApJ...413L.105P}, i.e. between $-18.5$ and $-19.7$.
We point out that this range, based on \cite{2010ApJS..190..418G}, is exactly equal to the 2$\sigma$ limits of our assumed luminosity function for normal events (Table~\ref{tab:MC}).
From Fig.~\ref{fig:IaIInAllowed}, we can see that the minimum brightness of a SN that could be classified as a Ia-CSM corresponds to the less luminous SN~Ia and the highest $f_V$ allowed (0.7). Similarly, the most luminous SN that could be classified as a Ia-CSM corresponds to the most luminous SN~Ia and the lowest  $f_V$ allowed (0.2).
We thus deduce that SNe~Ia-CSM can be found in the magnitude range $-19.5 > M_V > - 21.6$.
In addition, the inclusion of extinction would relax the lower luminosity limit allowing for fainter events.
These results are in excellent agreement with the findings of \cite{2013ApJS..207....3S}.
This shows that our simplified model can reproduce real events and that it has predictive power.

If we assume that strong CSM interaction is only related to the most luminous (91T- and 99aa-like) SNe~Ia (Sect.~\ref{sec:nobias}) and place an upper limit of e.g. $-19$ to $M_{SN}$, we will get a shorter magnitude range for Ia-CSM objects with the less luminous being at $-20$.
This could be in a possible tension with the Ia-CSM luminosities presented by \cite{2013ApJS..207....3S}.
We stress however, that this estimate ignores the effect of extinction that would certainly relax this tension \citep[e.g.][]{2012A&A...545L...7T}.
In addition, some of the faintest objects identified by \cite{2013ApJS..207....3S} could indeed belong to the region of Fig.~\ref{fig:IaIInAllowed} where `sub-typing is possible'. This is certainly the case for PTF11kx ($-19.3$), which was immediately recognized as 99aa-like \citep{2012Sci...337..942D}, and for PTF10htz ($-19.1$), which has clear features and therefore relatively high $f_V$ in the context of our model.
These are the two faintest objects in their PTF sample ($>-19.8$) and this can be attributed to lower continuum contamination.

\subsection{On the number of SNe~Ia disguised as SNe~IIn}
\label{sec:IaIIn}

It is also possible, based on Fig.~\ref{fig:IaIInAllowed}, 
to put tighter constraints on the fraction of SNe~IIn that might be of thermonuclear nature.
SNe~IIn (including IInS in our project) are bounded by $f_V < 0.3$.
Therefore, the faintest SN that can be classified as a IIn, under the assumption that it really contains a SN~Ia, is $M < -20.1$. 
This is a quite high luminosity limit and it becomes even higher if we consider a less conservative bound ($f_V < 0.2$).
This places strict limits on the number of SNe~Ia that are really misclassified as SNe~IIn. 
Especially after a careful inspection of SN~IIn spectra, such as the one performed by \cite{2013ApJS..207....3S},
it will be possible to miss SN~Ia signatures only among the brightest SNe~IIn.

\cite{2013ApJS..207....3S} searched for SNe~Ia-CSM in two different SN~IIn samples: the Berkeley SN database and among PTF discoveries.
The former sample is more diverse and contains SNe that have been mostly discovered by targeted surveys, i.e. by monitoring their host galaxies.
As pointed out by \cite{2013ApJS..207....3S}, very few of the SNe~IIn from this sample are among the brightest events, 
but most are found between $-17$ and $-19$ \citep[see also][]{2011MNRAS.412.1441L}.
This is also the luminosity range where most other targeted and well-studied SN~IIn samples lie \citep{Kiewe12,2013A&A...555A..10T}.
This suggests that it is not likely that there are many more SNe~Ia hidden below SNe~IIn among this sample and that the search of \cite{2013ApJS..207....3S} might have revealed all of them. 

We now focus on the PTF sample.
Although there might be numerous biases related to the selection of PTF discoveries for spectroscopic follow-up, the discovery method is independent of the host galaxy properties and we consider this sample to be more representative of the real population of SNe IIn (at least concerning the luminosity range).
Indeed, the sample spans a much wider range of absolute luminosities (from $-14$ to $-22$) and contains many more luminous events \citep[35/63 events are brighter than $-19$;][their figure 12]{2013ApJS..207....3S}. 
\cite{2013ApJS..207....3S} identified 7/63 of these objects to be hiding SNe~Ia in disguise. 
Based on the lower luminosity limit for a SN~IIn classification derived above, we conclude that  only SNe~IIn brighter than $M < -20$ can be hiding a thermonuclear SN, considerably lowering the number of candidates to 12 SNe. 
In fact, $f_V < 0.2$ would be more appropriate for a dedicated search as the one by \cite{2013ApJS..207....3S}, yielding a limit closer to $M < -20.5$. This would bring the maximum number of SNe~Ia in the PTF SN~IIn sample to 13/63 \cite[including the seven already discovered by][]{2013ApJS..207....3S}.
This number, however, is a strict upper limit as  a fraction of the high-luminosity events will most certainly be core-collapse SNe.

An interesting implication is to estimate whether the SN~IIn sample of \cite{2012MNRAS.424.1372A} could be contaminated by SNe~Ia-CSM. This could naturally explain their finding that SN~IIn locations do not trace the ongoing star formation in their hosts.
Their sample ($N=19$) definitely includes SNe that are known to be genuine core-collapse SNe~IIn (e.g. SNe~1994W, 1995N, 1996cr).
However, it also includes objects that are not well-studied and for which little information is known, apart from their classifications. 
We searched circulars and other literature \citep[e.g.][]{2011MNRAS.412.1441L,2013ApJ...763...42O} trying to place constraints on the absolute magnitudes of the SNe in their sample. 
Often, an accurate estimate is difficult because of the uncertain phase the reported magnitudes refer to (e.g. SNe~1996ae, 2002A, 2006am).
However, in most cases, the lower limits derived are too far (between $-14$ and $-16.5$) from the fiducial $-20.1$ threshold derived above, making it unlikely that these SNe could hide SNe~Ia.
We only found three SNe for which the approximate absolute magnitude limits are relatively high ($-18.5$ to $-18.8$): SNe~1999gb, 2000P, and 2000cl. 
The first two objects were examined by  \cite{2013ApJS..207....3S} who did not find any evidence for Ia signatures.
Without access to spectroscopy for the full sample, it is difficult to be certain about the nature of all objects. 
Nevertheless, from the brightness constraints we estimate that the thermonuclear contamination in the SN~IIn sample of \cite{2012MNRAS.424.1372A} cannot exceed one to two objects.
Therefore, the explanation for the weak correlation of SNe~IIn with star formation must be sought elsewhere.

\section{Stripped supernovae}    
\label{sec:Ic}

\subsection{Normal events}
\label{sec:IbcNormal}

\begin{figure}
\begin{center}
\includegraphics[width=\columnwidth]{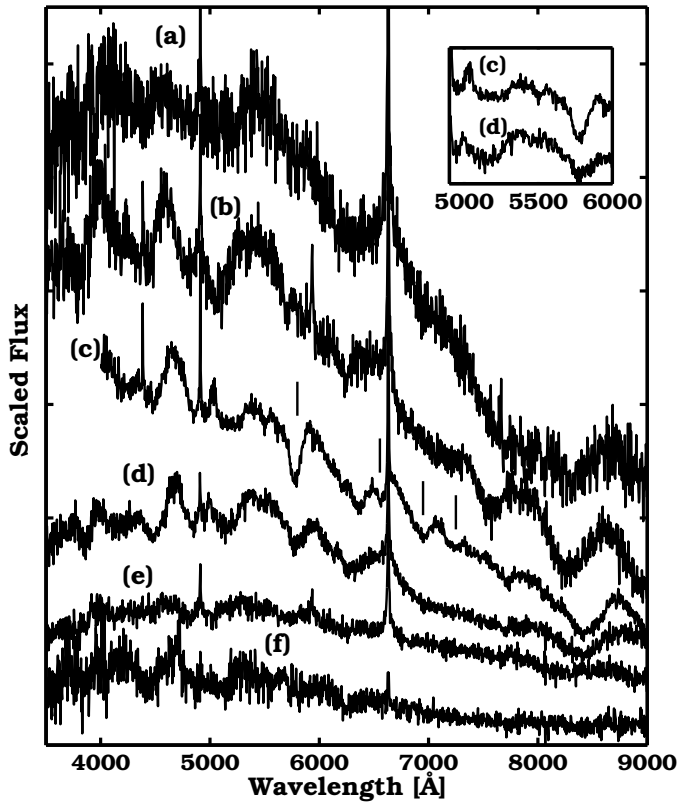}
\caption[]{From top to bottom: 
(a) a Ic-BL event (+20d) classified correctly; 
(b) a SN~Ic (+10d) classified correctly;
(c) the single SN~Ibc (+10d) that was misclassified as a 91T-like SN~Ia. \ion{He}{i} lines are indicated with the vertical bars; 
(d) a 91T-like event (+20d) misclassified as a SN~Ibc (one of two such events); 
(e) one out of the four SNe~Ibc  classified as Ia-CSM: +0d, $f_V = 0.17$ and S/N$=22$;
(f) one out of the four SNe~Ic  classified as 91T-like:  +0d, $f_V = 1.16$ and S/N$=6$;
The inset zooms in the region $5000-6000$ \AA\ for spectra (c) and (d): no significant difference is seen at these phases.}
\label{fig:Ibc}
\end{center}
\end{figure}

It is not clear if supernovae that have been completely stripped of their H envelope can interact strongly with a H-rich CSM \citep[although a connection with He-rich CSM has been established;][]{2008MNRAS.389..113P}.
In the case of a single massive progenitor, this type of a configuration would require a fine-tuning of the time of explosion with the time the last part of the H envelope was expelled. This seems unlikely. 
This condition is not required in the case of a binary system where the SN progenitor, likely a WR star, would explode within the wind created by its binary companion (similar to the picture we have for Ia-CSM SNe). 
A number of studies, using different diagnostics, have suggested that at least a fair fraction of SNe~Ibc come from binary progenitors
\citep[e.g.][]{2007PASP..119.1211F,smarttARAA,2011MNRAS.412.1522S,2011A&A...530A..95L}.
It is therefore theoretically possible that such events (i.e. SNe~Ibc-CSM) could exist in nature.

For the rest of the section, we will suppose that Ibc-CSM  events indeed exist and that they appear spectroscopically similar to our model spectra.
\cite{2006ApJ...653L.129B} have suggested that SN~2002ic might be such an object. 
By extension, this would imply that more (or all) Ia-CSM supernovae might be of core-collapse nature. 
There are several arguments against this view including the detailed spectroscopic differences between SNe~Ia and SNe~Ic, such as the \ion{S}{ii}  W-shaped feature at $\sim$5400~\AA\ (seen in SN~2002ic), or the existence of clear cases, such as PTF11kx \citep{2012Sci...337..942D}.
\cite{2014MNRAS.437L..51I} argued that SN~2012ca was another Ic-CSM event based on the likely detection of blue-shifted O, Mg, and C in late-time spectra. However, these line identifications were challenged by \cite{2014arXiv1408.6239F}, who propose that these lines can instead be due to coronal Fe emission, and thus favour a thermonuclear origin for this SN. 
Besides, \cite{2013ApJ...775L..43T} have demonstrated that even the presence of broad nebular [\ion{O}{i}] emission can be compatible with a thermonuclear explosion.

We offer two additional arguments that make the stripped core-collapse scenario unlikely. The first has to do with the observed magnitude distribution of Ia-CSM supernovae \citep{2013ApJS..207....3S}, which is very well reproduced by Type~Ia supernovae (Fig.~\ref{fig:IaIInAllowed}). In our experiment, SNe Ibc and Ic show the same critical $f_V$ ($\sim$0.3) for correct identification as SNe~Ia, and this means that a population of very bright SNe~Ic (between $-18.5$ and $-19.7$) would be required  to explain the Ia-CSM luminosity function.
This is very unlikely, as even SN~2004aw \citep{2006MNRAS.371.1459T} falls short of this magnitude range ($M = -18$).

The second argument relates to the limited number of misclassifications between simulated thermonuclear (even 91T-like) and stripped core-collapse SNe interacting with a CSM. 
In Sect.~\ref{sec:clas}, it was demonstrated that out of the 257 events that were classified as thermonuclear, only seven were SNe~Ibc of various sub-types.
The error was more substantial the other way around, showing that it is much more likely to mistake a 91T-like object for a Type~Ic (or Ibc) than the opposite. In other words, the chances are bigger that it is \cite{2006ApJ...653L.129B}, rather than \cite{2003Natur.424..651H} who have misclassified SN~2002ic  (based on spectroscopic template comparisons).

It is of general interest to discuss the individual misclassifications in more detail.
Figure~\ref{fig:Ibc} shows several of these cases including both the single SN~Ibc, which was classified as 91T-like, and a 91T-like object, which was classified as SN~Ibc (both at $f_V \sim 0.37$ and at 10-20 days past maximum). 
The similarities are indeed large and the inset shows that at these phases, the \ion{S}{ii} W-feature, which is present at earlier phases in thermonuclear SNe, cannot serve as a distinctive feature.
Nevertheless, the Nugent Ibc template demonstrates clear absorption features from \ion{He}{i} at 6678, 7150, and 7281 \AA\ that should have served as a discriminant. 
In addition, we show representative examples for SNe~Ic classified as 91T-like and stripped SNe classified as Ia-CSM.
Out of these eight cases in total, we observe that  five were at S/N $<20$ (three at S/N $<$10), while all others had particularly low $f_V$ ($<$0.17).
Therefore, as expected, at low S/N the danger of individual misclassifications increases.
However, none of the SNe that have been characterized (or even debated) as Ia-CSM in the literature were classified based on such poor data.

The arguments presented in this section were primarily statistical and apply better to the Ia-CSM sample as a whole.
However, we cannot exclude that individual events (e.g. a bright SN~Ic) can under certain circumstances (proper $f_V$, low S/N) be mistaken for an Ia-CSM.
We conclude that the majority of Ia-CSM SNe in the literature are indeed of thermonuclear nature. Contamination from core-collapse events, if existent, should be minor.

\begin{figure*}
\begin{center}
\includegraphics[width=\textwidth]{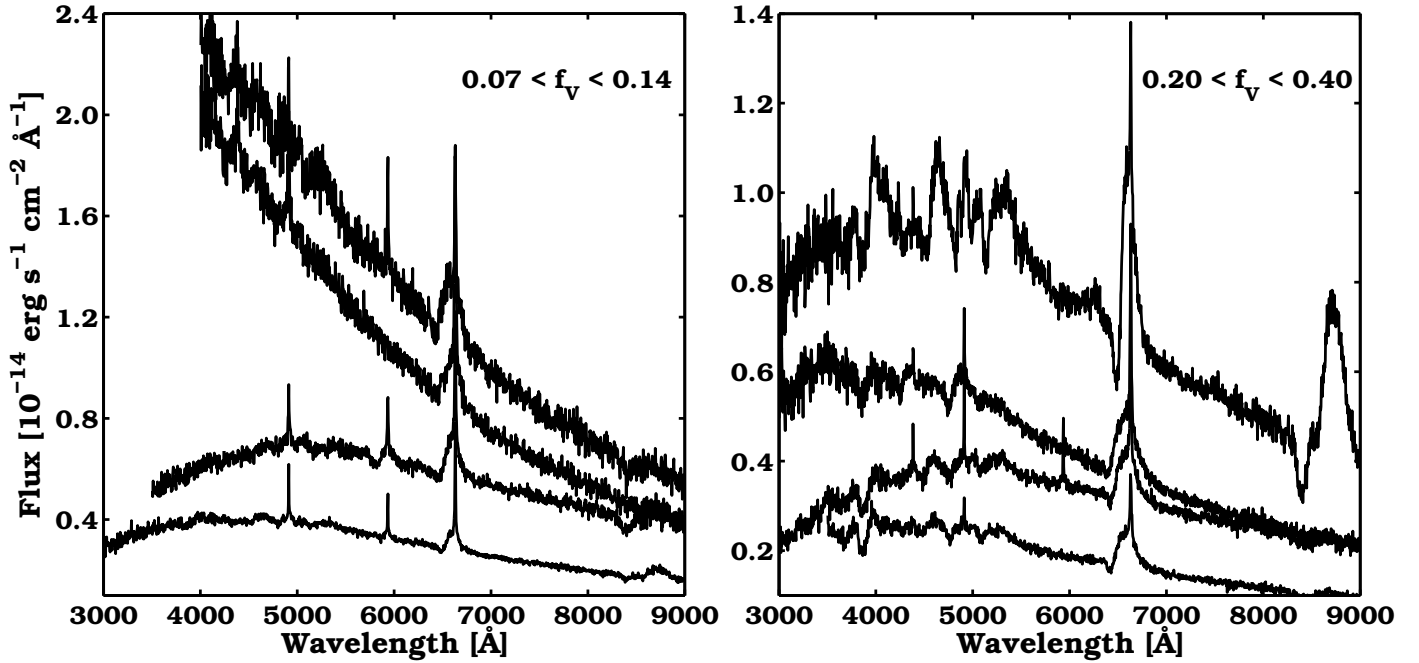}
\caption[]{Example spectra of SNe~IIP interacting with a CSM from our MC simulation. 
\textbf{Left:} Spectra  with a flux ratio $f_V$ between 0.07 and 0.14. 
 \textbf{Right:} Spectra  with $0.2 < f_V  < 0.4$. The spectra span a range of phases, $M_{SN}$, $A_V$, $T_{BB}$ , and other simulation parameters. The spectra to the right show more features and different H${\alpha}$ profiles than those to the left.
The spectra have not been scaled.
}
\label{fig:IIPseries}
\end{center}
\end{figure*}

\subsection{Broad-lined events}
\label{sec:Ic-BL}

In Sect.~\ref{sec:clas} it was shown that SNe~Ic-BL were relatively easy to identify in this experiment. 
This was because of their high luminosity and their broad-lined features. 
Figure~\ref{fig:Ibc} shows a characteristic example with $f_V = 0.4$.
However, all events below $f_V = 0.3$ were classified as IIn (or IInS). 
This means that a SN~IIn that can hide a SN~Ic-BL, such as the average assumed in this simulation, must be very bright ($M = -20.7$).
A fainter event, however, such as SN~2002ap \citep[$M = -17.5$;][]{2002ApJ...572L..61M}, could be successfully masked by strong CSM interaction resulting in a SN~IIn as faint as $M = -19$.
The fact that such events have not been recognized means that they must be intrinsically rare, if they exist at all.

\section{Type~II supernovae}     
\label{sec:II}

\begin{figure*}
\begin{center}
\includegraphics[width=\textwidth]{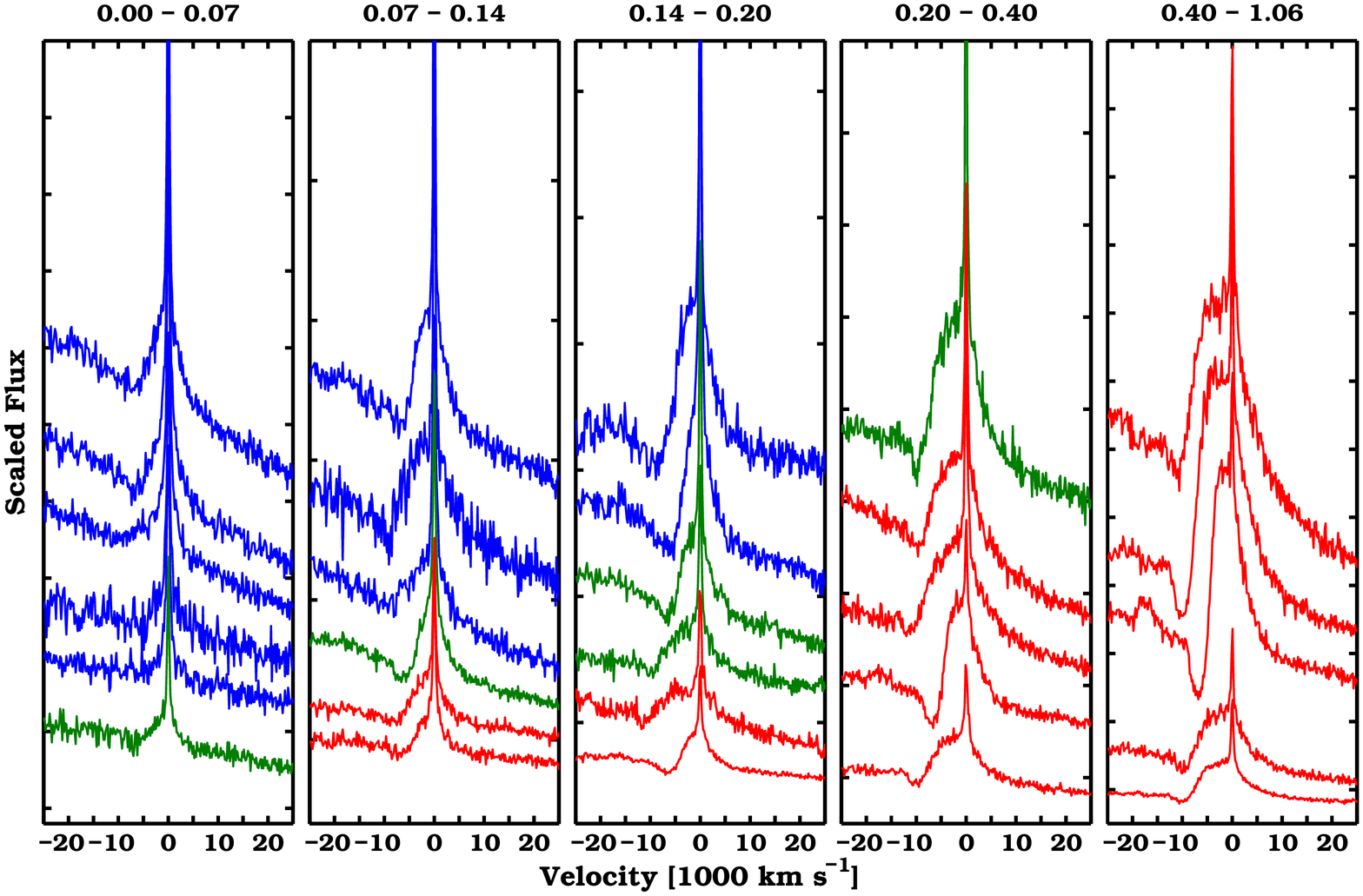}
\caption[]{Representative velocity profiles of the H${\alpha}$ lines of SNe~IIP-CSM for different values of the ratio $f_V$.
The range of each $f_V$ bin is shown above each panel.
The spectra are coloured depending on the classification they received in this project: blue for Type~IIn, green for Type~IInS, and red for Type~IIP.
As the relative SN contribution increases, the  H${\alpha}$ lines become broader, more asymmetric and the absorption component of the P-Cygni profile appears.
The flux of the spectra has been scaled for presentation purposes.
}
\label{fig:HaIIP}
\end{center}
\end{figure*}

It appears that it is possible to correctly classify SNe~IIP at lower flux ratios than what was done for other SN Types, including SNe~Ia. We find that SNe with an underlying IIP template were given IIn classifications if they had $f_V$ between 0.01-0.08, IInS classifications for flux ratios between 0.02-0.14, and IIP classifications for flux ratios between 0.11-0.44 (excluding 16\% outliers from both sides). So the critical limit seems to be pushed more towards $f_V \sim 0.15$ (seven times more continuum than SN emission!), as opposed to 0.2-0.3 for thermonuclear SNe. 
Similar to the other SN types, we did not find any significant dependence on any parameter that is not degenerate with $f_V$, such as extinction, S/N, FWHM, etc.
We conclude that the fact that it is easier to identify Type~II SNe at lower $f_V$ is due to their spectra having strong and broad emission lines (H$\alpha$), rather than being dominated by absorption. The ranges obtained for $f_R$ are very similar to those for $f_V$, showing that the differences are indeed due to spectral features and not to broad-band properties.

Figure~\ref{fig:IIPseries} shows some selected example spectra in two different regimes of the relative SN contribution. At low $f_V$ the spectra are relatively featureless and the H$\alpha$ profiles are more symmetric. At higher $f_V$ the Type~II nature of the spectra becomes more pronounced.
The H$\alpha$ profiles are examined in more detail in Fig.~\ref{fig:HaIIP}. Indeed, we confirm that the profiles become broader, less symmetric, and extend to higher velocities as we move to higher $f_V$. In addition, the absorption component of the P-Cygni profile becomes more prominent. 
It is not surprising that this leads to more SNe being classified as IIP.
We observe, however, many similar profiles that are classified in different ways (IIn, IInS or IIP), especially in the intermediate regimes, implying that the human factor is also important in this case. 

Many of the spectra shown in Fig.~\ref{fig:IIPseries} are very similar to real SNe that have either been classified as SNe~IIn or SNe~IIP in the literature. They span a wide range of luminosities and spectral appearances, making it possible that these events have exact matches in nature and that CSM interaction plays an important role in Type~II explosions for a wide range of the ratio $f_V$. 
In fact, many real objects lie in an intermediate regime \cite[e.g.][]{2010ApJ...715..541A,2012MNRAS.422.1122I}, including objects that transition from one class to another \citep[e.g.][]{2013A&A...555A.142I}.

\section{What is below the faintest SNe~IIn?}     
\label{sec:faint}

It is now  possible to place constraints on the nature of a SN~IIn, depending on its luminosity.
This is based on the fact that the explosion that has caused the ejecta-CSM interaction, if an explosion at all, cannot be brighter than a certain limit. Otherwise, it would leave clear spectral signatures that would be possible to identify.

\begin{figure}
\begin{center}
\includegraphics[width=\columnwidth]{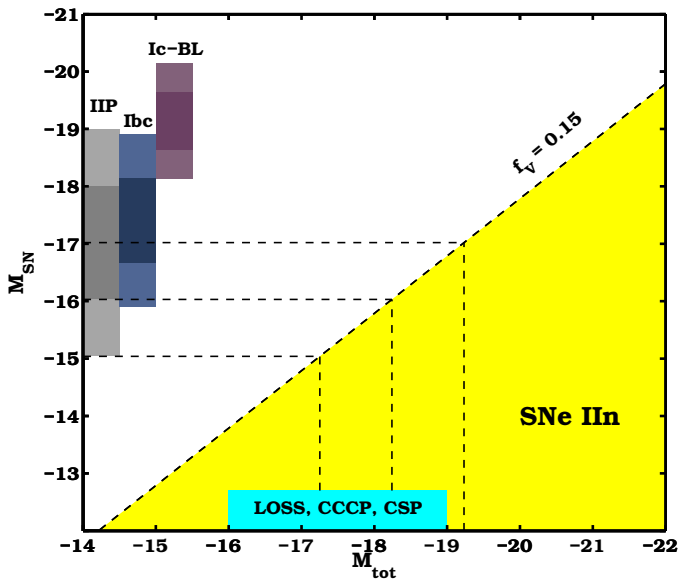}
\caption[]{Maximum luminosity of a SN below strong CSM interaction as a function of the resulting total luminosity and the ratio $f_V$.  
SNe that are classified as IIn are found in the region $f_V < 0.15$ (yellow-coloured area; limit best suited for IIP-CSM) and there is a corresponding upper limit in the maximum luminosity of the underlying transient. The luminosity range for most well-studied samples of SNe~IIn is indicated by the cyan colour. LOSS, CCCP, and CSP stand for the samples of  \cite{2011MNRAS.412.1441L}, \cite{Kiewe12} and \cite{2013A&A...555A..10T}, respectively.
1$\sigma$ and 2$\sigma$ contours for the luminosity functions of the core-collapse SNe adopted here are shown as vertical bars.
A few horizontal and vertical dashed lines are drawn to guide the eye. 
It is shown that the faintest ($M \gtrsim  -17.2$) SNe~IIn cannot hide a SN~IIP that is brighter than $M=-15$.}
\label{fig:IInAllowed}
\end{center}
\end{figure}

In Fig.~\ref{fig:IInAllowed} we have plotted the absolute magnitude of the underlying SN as a function of the total absolute magnitude of a CSM interacting SN, i.e the opposite of Fig.~\ref{fig:IaIInAllowed}. The objects that are classified as IIn are found in $f_V < 0.15$ (yellow area).
This value is optimized for SNe~IIP, while it was shown in the previous sections that it is larger ($\sim$0.2--0.3) for SNe~Ia and Ibc.
From this graph, it is possible to deduce the maximum allowed SN luminosity that can be hidden by the CSM interaction. 

Most well-studied SNe~IIn are found in the region  $ -16 > M > -19$ \citep{2011MNRAS.412.1441L,Kiewe12,2013A&A...555A..10T}.
The brightest of these events are consistent with harbouring SNe~II that have a luminosity function of $M_{V} = -17.02 \pm 0.99$ \citep{2014ApJ...786...67A}.
However, this becomes increasingly difficult for the faintest SNe~IIn.
Faint SNe~IIP have $M \sim -15$ \citep[e.g.][]{2004MNRAS.347...74P} and cannot be hidden below SNe~IIn fainter than $M=-17.2$.
Even SNe~IIn as bright as $M=-18$ can only efficiently hide events in the  lower 2-5\% of the SN~II luminosity function of \cite{2014ApJ...786...67A}.
Therefore, if caused by core-collapse, these events must be due to extremely faint explosions \cite[e.g.][]{2007Natur.449E...1P,2009Natur.459..674V}.
However, these events are both very rare and their core-collapse nature has been debated \citep{2007Natur.447..458K,2009AJ....138..376F}.
Explosions like SN~2008S, that have been proposed to be electron-capture SNe, are also in the right magnitude range \citep{2008ApJ...681L...9P,2009MNRAS.398.1041B,2009ApJ...705.1364T}.
Alternatively, faint SNe~IIn are not due to SN explosions but to stellar eruptions that do not destroy the progenitor star  \citep[SN impostors;][]{2000PASP..112.1532V,2006MNRAS.369..390M}.
Of course, this conclusion is not new, but we have derived it here in an independent and quantitative way.

\section{On the adequacy of the adopted model}  
\label{sec:mod}

The model behind our MC simulation is simplified: 
the emission from the SN and CSM interaction are treated independently and it is assumed that they can be added to obtain a composite spectrum.
It is conceivable, however, that radiative transfer processes in the CSM \citep[e.g.][]{2001MNRAS.326.1448C} could alter the underlying SN spectrum and thus affect the conclusions of this study.
In this section we provide a number of arguments as to why our approximation is reasonable to a first order and why there are at least a subset of conditions and geometrical configurations where it should be valid.

First, we show that even a CSM resulting from a spherically symmetric and steady wind 
will be optically thin for most reasonable assumptions.
When the shock wave is at $R_\mathrm{s}$, the swept-up CSM mass is
$\dot{M}R_\mathrm{s}/v_\mathrm{w}$ where $\dot{M}$ is the mass-loss rate and $v_\mathrm{w}$ is the wind velocity.
We assume that the swept-up mass is in a dense shell whose  width $\Delta R$ is much smaller than $R_\mathrm{s}$
$(\Delta R\ll R_\mathrm{s})$, which is typically true for SNe IIn because of the radiative cooling.
Assuming, for simplicity, that  the density $\rho_\mathrm{shell}$ in the shell is constant, 
the optical depth $\tau_\mathrm{shell}$ of the shocked shell becomes:

\begin{equation}
\tau_\mathrm{shell}=\kappa_\mathrm{shell}\Delta R\rho_\mathrm{shell}
=\frac{\kappa_\mathrm{shell}\dot{M}}{4\pi R_\mathrm{s}v_\mathrm{w}}
,\end{equation}

where $\kappa_\mathrm{shell}$ is the opacity of the shocked shell.
Therefore, for the typical mass-loss rates of SNe~IIn \citep[$\sim10^{-4}-10^{-2}~M_\odot~\mathrm{yr^{-1}}$, e.g.][]{Kiewe12,2013A&A...555A..10T},
$\tau_\mathrm{shell} < 1$ at the typical BB radii used in this paper ($\sim10^{15}$~cm)
and the optical radiation from the SN is not expected to be affected by the dense shell.
In reality, there should also be contributions from the shocked SN ejecta to the optical depth.
However, this will only introduce a small correction factor because the shocked ejecta mass is comparable to the shocked
CSM mass.  
To summarize, the assumption that the emission from the SN and the CSM is separated and additive is a good approximation for our models.

\begin{figure}
\centering
\includegraphics[width=\columnwidth]{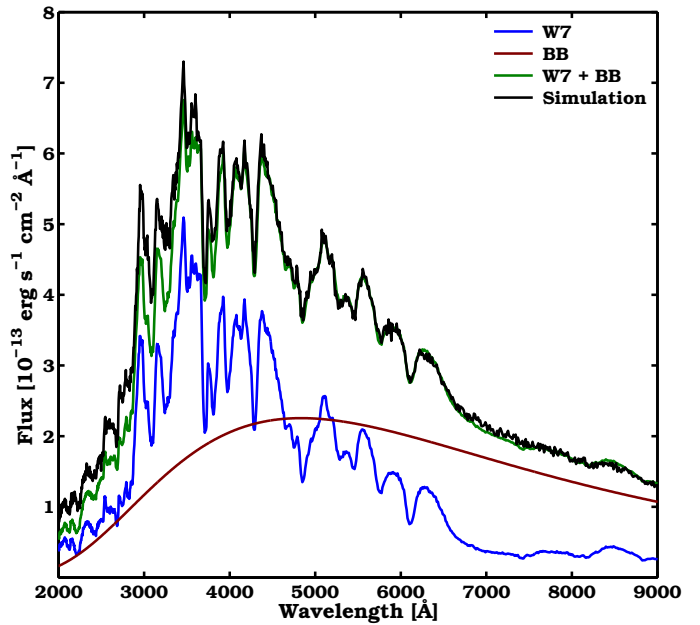}
\caption{
Results from our radiative transfer simulation, verifying that the SN spectrum is not significantly altered by the CSM interaction.
This `snapshot' simulation is for a SN~Ia at maximum light, with the W7 model of \cite{1984ApJ...286..644N} (blue) and a BB component with temperature 6000~K (wine). In the optical wavelengths, their sum (green) is not significantly different than the fully simulated spectrum (black), for the (relatively strong) mass-loss parameters assumed here. The simulations refer to a distance of 10 Mpc.
}
\label{fig:W7comp}
\end{figure}

To confirm this, we have performed radiation transfer simulations based on a toy model of the interaction. The simulations have been performed using  a newly developed multi-dimensional and multi-frequency code (K.~Maeda, in prep.), 
which adopts prescriptions similar to \cite{2006ApJ...651..366K} under the time-dependent Monte-Carlo approach and LTE approximation. 
As an input, we have used the W7 model of a SN~Ia \citep{1984ApJ...286..644N}. 
The specific run assumes spherical symmetry, although the code can handle an arbitrary geometry.
To mimic the CSM interaction radiation field, photons are  created following a Planck source function with a temperature of 6000~K, 
with a total luminosity of $2 \times 10^{43}$ erg, at a shell with velocity of $18000$ km s$^{-1}$. This photon source is added to 
the simulation together with the photons created by radioactive decays. 
The luminosity and velocity are adopted to roughly mimic the interaction from a large mass loss with $10^{-2} M_{\odot}$ yr$^{-1}$, for $v_{\rm w} 
= 100$ km s$^{-1}$ \citep[][]{1982ApJ...258..790C,2014MNRAS.439.2917M}
Figure~\ref{fig:W7comp} compares the sum of the W7 and BB spectrum around peak, 
i.e. equivalent to our simple MC additive approach,
and the spectrum of the W7 model irradiated from outward as described above. 
It is shown that the optical spectrum remains largely unaffected, even for  the relatively strong mass-loss parameters adopted.
There are indications that there might be larger deviations in the UV but these will not affect our conclusions concerning line-identification and classification.
A detailed discussion is beyond the scope of this paper and these results will be discussed elsewhere.

\begin{figure}
\begin{center}
\includegraphics[width=\columnwidth]{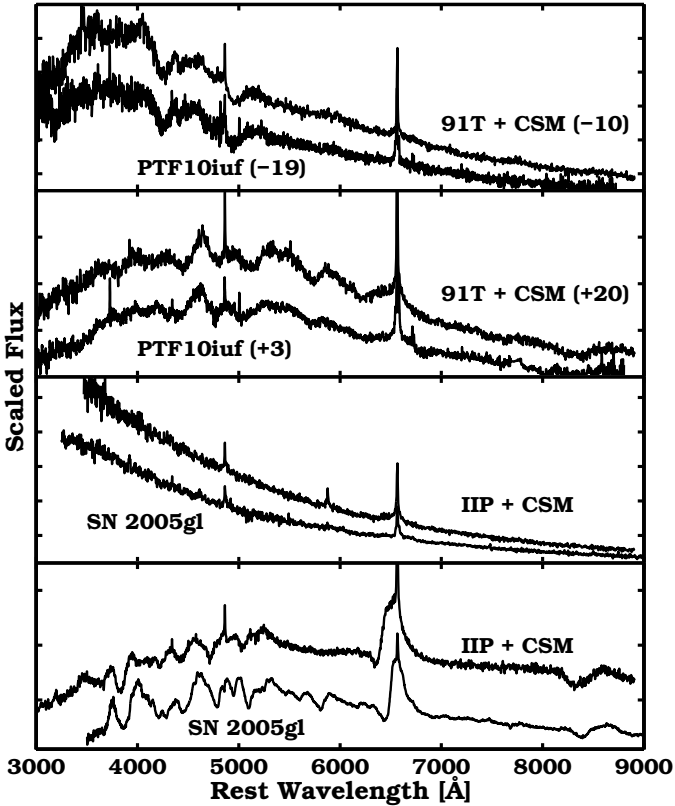}
\caption[]{Example spectra from our MC simulation compared to real SNe~IIn and Ia-CSM.
No effort has been made to fit the data, but the spectra were selected from our simulated sample based on their similarity to the observed spectra. 
The top two panels show a comparison with the Ia-CSM PTF10iuf that has good constraints on the spectroscopy epochs \citep{2013ApJS..207....3S}. The early spectrum (-19d) is well fit with a 91T-like SN at -10 days with $f_V = 0.67$ and $A_V = 0.71$.
Normal SNe~Ia do not give an acceptable match.
The spectrum at +3d continues being consistent with a 91T-like object. The phase of the underlying template is +20d and $f_V = 0.27$.
The bottom panels show that simulated IIP-CSM SNe can provide good matches to several SNe~IIn, such as SN~2005gl
\citep{2007ApJ...656..372G}. The early, hot, spectrum of SN~2005gl is well reproduced by a SN~IIP with $f_V = 0.04$ (T$\sim$13000 K).
At this low $f_V$ the spectral shape is not sensitive to the underlying type and any template could provide a hot SN~IIn spectrum. 
At later times, however, when SN~2005gl develops clear Balmer lines, a SN~IIP with $f_V = 0.51$ provides a remarkably good match.
}
\label{fig:real_data_comp}
\end{center}
\end{figure}

In addition to this idealized spherically symmetric case, there are a number of realistic geometrical configurations that are compatible with our assumption. In one of them, the CSM still fully covers the SN photosphere but it is clumpy. In this case, there are many lines of sight that let the SN light escape through the CSM unaffected. At the same time, continuum photons are scattered from the shocked high density regions towards our direction, justifying this additive process.  

A similar situation can occur if the CSM only partially covers the SN photosphere. This can happen if the CSM is e.g. located in a torus that is inclined with respect to our line of sight.  It is stressed that the ideas of a clumpy or asymmetrical CSM  have been extensively proposed in the literature to explain SN~IIn observables \citep{1994MNRAS.268..173C,2002ApJ...572..350F,Smith06tf,2011MNRAS.412.1639D,2012Sci...337..942D,2014AJ....147...23L}.
It is reasonable to believe that these ideas are compatible with observations \citep{2007AJ....134..846S} and realistic theories of mass loss  \citep[][and references therein]{2008A&ARv..16..209P}.

Finally, the composite spectra from our simplified model are very similar to real data (Fig.~\ref{fig:real_data_comp}).
Our method presents the obvious advantage that large statistical samples can easily be  constructed.
Detailed radiative transfer modelling will be  valuable to verify the validity of the main assumptions and conclusions of this study, but it will be computationally impossible to explore large regions of the parameter space, as done with the present MC approach.

\section{Summary and conclusions}       
\label{sec:conc}

In this work, we have simulated spectra of supernovae interacting strongly with a CSM.
This was done by adding the fluxes of SN template spectra, a BB continuum, and an emission line spectrum.
A number of parameters, such as the SN type, luminosity, and epoch, the BB temperature and radius, the strength and width of the emission line components, the extinction and the S/N, were varied in  an MC way and 823 spectra were finally constructed. 
These spectra were distributed to ten different observers for classification. For the needs of this project, we have introduced the class of IInS supernovae (where the S stands for structure). The idea behind this is to distinguish between featureless SNe~IIn and those that present some weak structure over the continuum.
We provide the results of this classification procedure and discuss how the CSM interaction affected the typing of the SNe. 

Despite that 91T-like SNe~Ia were on average 0.4 mag brighter than normal events, we found no luminosity bias suggesting 
that their identification would be easier when interacting with (and contaminated by) a CSM.
Combined with the observed association of 91T-like events with the observed SNe~Ia-CSM, and their small fraction among SNe~Ia, we suggest that this sub-class of luminous thermonuclear explosions results from single degenerate systems.
This is reinforced by the fact that both 91T-like and Ia-CSM SNe are found in late-type, star-forming host galaxies.

We have defined and parametrized our discussion with respect to the ratio of the contributing SN and BB fluxes ($f_V$ when measured in the $V$-band). We have convincingly shown that this is the single important parameter  significantly affecting the composite spectrum's appearance and classification. The classification is not affected by any other parameter that is not degenerate with $f_V$, including the extinction, the S/N, the strength and the width of the emission lines and the presence of \ion{He}{i}.
For thermonuclear SNe, it was shown that a correct determination of their nature was possible when the ratio $f_V$ was greater than $\sim$0.2-0.3. This means that correct classification of a SN is still possible, as long as the continuum radiation from the CSM interaction is not more than (about) five times the emission from the SN itself. Since the classification did not depend significantly on the narrow emission lines, we suggest that this result can be generalized  for  host galaxy contamination, i.e. another type of  contamination, at least for star-forming galaxies and red-wards of the 4000 \AA\ break, where the continuum is relatively smooth.

We showed that SNe~Ia get progressively classified as Ia-CSM (where no further sub-typing is possible), IInS and IIn as the ratio $f_V$ decreases, and we determined the boundaries where these transitions occur. Using this result, and assuming a standard luminosity function for SNe~Ia, we were able to predict that SNe~Ia-CSM occur in the magnitude range  $-19.5 > M_V > - 21.6$ (un-extincted). This is in very good agreement with the observational work of \cite{2013ApJS..207....3S}. Furthermore, we were able to place a lower limit of 
$M_V < -20.1$ in the luminosity of any classified SN~IIn that might be hiding a SN~Ia. This places tighter constraints than before on the possible contribution of thermonuclear explosions in the observed zoo of SNe~IIn. In particular, it is unlikely that the SN~IIn sample of \cite{2012MNRAS.424.1372A} is significantly contaminated by SNe~Ia.

We have also examined the possibility of stripped SNe interacting strongly with a CSM and whether such explosions could be misidentified as SNe~Ia.
Although the spectroscopic similarities between these objects are strong, we have found that the amount of misclassifications is small (especially from Ia to Ibc sub-types). In addition, a population of very bright stripped SNe would be required to reproduce the observed magnitude distribution of SNe~Ia-CSM.
We therefore conclude that the population of SNe~Ia-CSM is primarily, if not uniquely, due to thermonuclear explosions.
The SNe~Ic-BL were easy to identify in our experiment because of their characteristic broad features and high luminosities. 
The fact that no such SN has been identified in nature to strongly interact with a CSM suggests that these events are intrinsically rare, if they exist at all.

It was demonstrated that Type~II SNe can be correctly recognized at lower flux ratios, down to $f_V \sim 0.15$. This was attributed to the strong emission components of the Balmer lines.  We show how the H$\alpha$ profiles change with increasing $f_V$ and the classifications progressively shift from SNe~IIn to SNe~IIP. The lines become broader and less symmetric, they extend to higher velocities, and the absorption component of the P-Cygni profile starts appearing.

We placed constraints on the nature of SNe~IIn, focusing in particular on the faintest events. While it is shown that a SN~IIn of $M_V=-19$ can efficiently  hide a normal SN~IIP, it becomes difficult for events that are fainter than $M_V = - 17.2$ to accommodate even the faintest SN~IIP ($M_V \sim -15$). Therefore, events fainter than this limit might either be due to extremely faint explosions  or they are simply stellar eruptions not destroying the progenitor star (SN impostors).

Despite the fact that our model is simplified and that it ignores radiative transfer, we have shown that it is a good first order approximation for reasonable assumptions affecting the optical depth of the CSM. We have checked against a more advanced model, including radiative transfer, and have verified the adequacy of our method, for a subset of the parameter space. 
In addition, we have argued that there are a number of geometrical configurations, such as a clumpy CSM or a CSM not completely covering the SN photosphere, which would naturally lead to spectra, such as those we have simulated.

\begin{acknowledgements}

We thank Joe Anderson for sharing with us the results on the absolute magnitude distribution of Type~II SNe prior to publication. 
We also thank Ariel Goobar and Rahman Amanullah for discussions, and Jens Hjorth for comments on the manuscript. 
GL was supported by the Swedish Research Council through grant No. 623-2011-7117 during the time this work was carried out. 
EYH and MDS gratefully acknowledge generous support provided by the Danish Agency for Science and Technology and Innovation realized through a Sapere Aude Level 2 grant.
The work by KM is supported by JSPS Grant-in-Aid for Scientific Research (23740141, 26800100).
TJM is supported by Japan Society for the Promotion of Science Postdoctoral Fellowships for Research Abroad (26á51).
JMS is supported by an NSF Astronomy and Astrophysics Postdoctoral Fellowship under award AST-1302771.
This research is supported by World Premier International Research Center Initiative (WPI Initiative), MEXT, Japan.
The Oskar Klein Centre is funded by the Swedish Research Council.
The Dark Cosmology Centre is funded by the Danish National Research Foundation.

\end{acknowledgements}


\bibliographystyle{aa}  
\bibliography{IaCSM.bib}

\end{document}